# DIGITAL CIRCUITS IMPLEMENTATION ON RPGA SIMULATOR

A

**Dissertation**

Submitted in partial fulfillment

for the award of degree of

*Master of Technology*

**(with specialization in Computer Science)**

*in Department of Computer Science and Engineering*

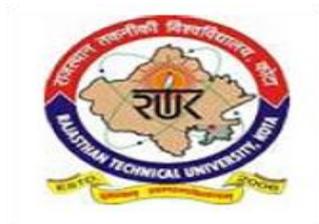

| **Supervisor** | **Submitted By:** |
|---|---|
| Dr SC Jain | Pankaj Kumar Israni |
| Professor, CSE | Enrollment No.: |
| | 11E2UCCSM4XP608 |

**Department of Computer Science and Engineering**

**University College of Engineering**

Rajasthan Technical University

Kota ( Rajasthan)

**JULY 2013**

# Candidate's Declaration

I hereby declare that the work, which is being presented in the Dissertation, entitled "**Digital Circuits Implementation On RPGA Simulator**" in partial fulfillment of "**Master of Technology**" with specialization in Computer Science and Engineering, submitted to the **Department of Computer Science & Engineering**, University College of Engineering, Rajasthan Technical University, Kota is a record of my own investigations carried under the guidance of **Dr. S.C. Jain Professor**, Computer Science, University College of Engineering, RTU Kota. I have not submitted the matter presented in this Dissertation Report any where for the award of any other degree.

**Pankaj Kumar Israni**

Computer Science Engineering

Enrollment No. : 11E2UCCSM4XP608

University College of Engineering, RTU, Kota(Raj)

**Counter Signed by**

Supervisor

**Dr. S.C. Jain**

Professor,

Department of Computer Science & Engineering,

University College of Engineering,

Kota (Rajasthan)

i

# Certificate

This is to certify that this Dissertation entitled "**Digital Circuits Implementation On RPGA Simulator**" has been successfully carried out by **Pankaj Kumar Israni** (Enrollment No.: 11E2UCCSM4XP608), under my supervision and guidance, in partial fulfillment of the requirement for the award of **Master of Technology** Degree in **Computer Science & Engineering** from **University College of Engineering**, Rajasthan Technical University, Kota for the year 2011-2013.

**Dr. S.C. Jain**

Professor,

Department of Computer Science & Engineering,

University College of Engineering,

Kota (Rajasthan)

# Acknowledgments

It is matter of great pleasure for me to submit this report on dissertation entitled "**Digital Circuits Implementation On RPGA Simulator**", as a part of curriculum for award of "Master in Technology with specialization in Computer Science & Engineering" degree of Rajasthan Technical University, Kota.

I am thankful to my Dissertation guide **Dr. S.C. Jain**, Professor in Department of Computer Science for his constant encouragement, able guidance and for giving me a new platform to build by career by giving me a chance to learn different fields of this technology. I am also thankful to **Mr. C.P. Gupta**, Asso.Prof. & Head of Computer Science Department for their valuable support.

I would like to acknowledge my thanks to entire faculty and supporting staff of Computer Engineering Department in general and particularly for their help, directly or indirectly during my Dissertation work.

I express my deep sense of reverence to my parents and family members for their unconditional support, patience and encouragement.

DATE                                                                                          Pankaj Kumar Israni

# CONTENTS

# List of Figures

# List of Tables

# List of Algorithms

# ABSTRACT

Reversible Computing is an emerging research area and is a promising technology for the next generation of computers. This is a new alternative for Low Power design solutions, Improved Reversibility, High Packaging density. Many researcher worked on irreversible computing have shown that there is high energy loss because there is information loss in irreversible computing. So, irreversible computing can not be used to build low power circuits and future computer. So, Reversible logic circuits play a very important role in design of low power circuits of a future computer. So, this technology is very useful in building important circuits related to advanced computing, low power CMOS design, optical computing, Quantum computation and nanotechnology.

For Reversible computing, Target technology is yet to become available. Adequate tools are not yet developed for reversible technology. Simulation is still under development. Classical logic synthesis methods and simulation tools can not be used in reversible computing. Because these work on irreversible logic blocks but in reversible computing reversible logic blocks are used for design and implementation. So, we need a simulation platform to analysis and development in this area. So, we worked on this area and developed a GUI based simulator.

In the Dissertation work, we undertook simulation of RPGA, reversible alternative to FPGA. We developed a RPGA simulator in our project. RPGA simulator combines the technology of PLD/FPGA/SYSTOLIC array. Our RPGA simulator, implements a given symmetric reversible Circuit on a RPGA structure. We also worked on RPGA structure gates (picton and kerntopf gates) to develop a better RPGA structure. Stepwise execution on RPGA structure is also performed in RPGA simulator. We also designed new algorithms for Truth Table generation and symmetry analysis. The dissertation work aims to develop entire simulator is GUI based and easy to learn that makes it user friendly. User can easily view all simulation results in GUI of our RPGA simulator.

# INTRODUCTION

A computation is reversible if the output contains sufficient information to reconstruct the input, no input information is erased [Toffoli 1980]. Reversible computing is different from the conventional computing. In the conventional computing, logic elements are normally irreversible in nature. In 1961, according to Landauer's principle [1] computing with irreversible logic results in energy dissipation. This is because erasure of each bit of information dissipates at least KTln2 joules of energy where k is Boltzamann's constant and T is the absolute temperature at which the operation is performed. By 2020 this energy loss will become a substantial part of energy dissipation, if Moore's law continues to be in effect. This particular problem of VLSI designing was realized by Bennet and others in 1970s. In 1973 Bennett [2] had shown that kTln2 energy dissipation would not occur, if a computation is carried out in a reversible way. According to Launder [1] and Bennet [2] losing information is equivalent to losing power consequently reversible computation is the best alternative for all future low power technologies. The subject of reversible computing is quite developed now. A pioneering step in this development was due to Toffoli who introduced the notion of universal reversible gates in 1980 which paved way for reversible computation.

Reversible logic offers low power (ideally zero) design solutions, since the input vector of reversible circuit can be uniquely recovered from the output vector. According to Bennet [2] zero energy dissipation would be possible only if the network consists of reversible gates. Thus, reversible logic is likely to be in demand in high speed power aware circuits. It is Suitable for Quantum/reversible/optical computing. It is the building block of reversible computation and improved the security, heat dissipation over irreversible logic circuits. The most prominent application of reversible logic lies in quantum computers as they must be built from reversible logic. Researcher also proving that next generation computers are quantum computers that will be constructed using reversible logic circuits. Researcher states every future technology will have to use reversible gates in order to reduce power. This has led many people to pursue research in the area of reversible logic.

## 1.1 Conventional Computing System and its Limitations

Conventional Computing system has reached to its upper limits of Design complexity, processing power, memory, energy dissipation, density and heat dissipation [3]. Therefore there is need of new alternative computing technology, which can overcome all these conventional computation problems. Moore's Low describes a long-term trend in the history of computing hardware, in which the number of transistors that can be placed inexpensively on an integrated circuit has doubled approximately every two years [3]. Conventional computing is based on silicon chips. Although silicon based computers have renovated the whole world from the start of computer age(1957 to till now). Since its invention, the whole advancement in silicon based technology is according to the Moore's Law and is achieved by doubling the processing speed and memory capacity at a very high rate. These tasks of high speed and memory capacity have achieved by reducing the size of components on chip and placing large number of transistors on IC chips [4]. On the other hand, due to this continually doubling the numbers of transistors on these chips, also reducing the size of these chips and increasing density will cause some serious physical problems in stable computing, in the next few years. Moreover, this technology is going towards its decay.

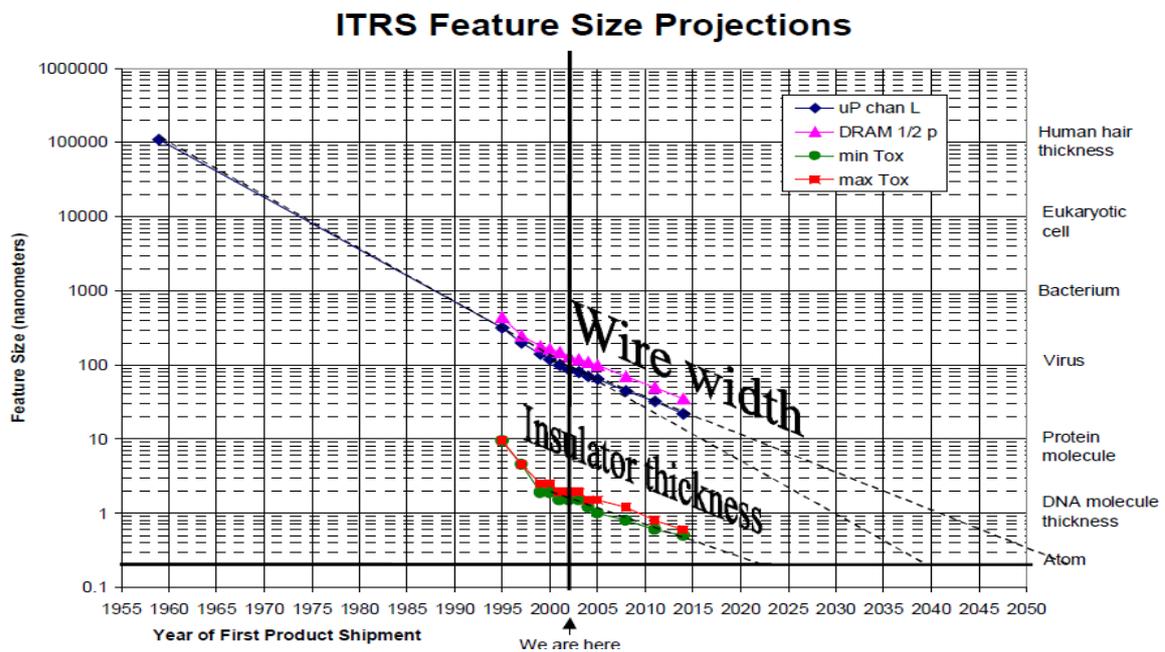

**Fig. 1.1. Trends for minimum feature size in semiconductor technology**

Fig. 1.1 shows Trends for minimum feature size in semiconductor technology. The data in the middle are taken from the 1999 edition of the International Technology Roadmap for semiconductors. The point at left represents the first planar transister fabricated in 1959. ITRS targets have historically always turned out to be conservative, so far. From this data, wire widths, which correspond most directly to overall transistor size, would be expected to reach near-atomic size by about 2040-2050. Conventional computing system have following limitations:

### 1.1.1 Speed and density

We know the phrase 'processing power of a computer' refers how fast a computer can execute an instruction that is given to a computer. All conventional computers achieved high speed by reducing the distance between different IC chip components and shrinking the size of processing chip and transistors on it, so that instruction take less time to move from one component to another during execution. For this, designers have to package more and more transistors (closely with each other) on IC chip and making chip denser and denser [3].

### 1.1.2 Design Complexity

As mentioned above, due to shrinking of IC chip size with doubling the transistors on it, chip manufacturing becomes complex and if this trend of shrinking and doubling continues, design complexity of chip would get tremendous increase, it will not be feasible for designers to develop new chips in the near future. Further extension of Moore's Law will face new more complex-design challenges [3].

### 1.1.3 Non - Recurring and High Cost

Investment over every new IC chip design will increase. As every new chip is more complex from the preceding one, therefore reducing size will not only add complexity to the design but it will also increase the overall cost per design from one chip to another [3].

### 1.1.4 Power Consumption and Heat Dissipation

Power consumption and heat dissipation is large obstacle for further advancement in conventional computing [5].

for every generation. This power consumtion also inverts the rare position effects of advancement in the number of transistors on silicon chip. This large amount of power consumption boosts up the heat generation, increasing danger that transistors interfere with each other. Heat dissipation, Power consumption are major limitation with which traditional computing are suffering. Therefore, there is need of searching for new alternative, which can solve all computational problems cited before [6].

## 1.2  Reversible Computation

Reversible computation [7] in a system can be performed only when the system comprises of reversible gates. In Computer Science, reversible transformations have been popularized by the Rubik's cube and sliding-tile puzzles, which fueled the development of new algorithms, such as iterative deepening A*-search [korf 1999]. Prior to that, reversible computing [8] was proposed to minimize energy loss due to the erasure and duplication of information .

### 1.2.1  Reversible Logic

First we give the definition of a reversible function. A Boolean function is reversible if a.) It has the same number of inputs and outputs.

b.) There is a one-to-one correspondence between the input vectors and the output vectors [9].

As we know that many traditional Boolean functions, such as AND, OR, XOR, NAND, are irreversible since all of them have more than one input but only one output. Also there is no one-to-one matching between the input vectors and the output vectors. However, the NOT function is reversible since it satisfies the definition above: it has one input and one output, and for each input vector, there is a distinct output vector corresponding to it so that the computation could be done from input to output, and inversely, from output to input as well [9].

Table 1.1 lists the truth table of a 2-input reversible gate called Controlled-NOT gate (or CNOT, Feynman gate). CNOT has two inputs (A and B) and two outputs (P and Q). It is shown in below truth table that there is a unique output vector corresponding to each input vector.

**Table 1.1. Truth table of Reversible Cnot gate**

| A | B | P | Q |
|---|---|---|---|
| 0 | 0 | 0 | 0 |
| 0 | 1 | 0 | 1 |
| 1 | 0 | 1 | 1 |
| 1 | 1 | 1 | 0 |

### 1.2.2 Reversible Circuits

A combinational reversible circuit is an acyclic combinational logic circuit in which all gates are reversible, and are interconnected without explicit fanouts and loops. Reversible circuits comprise a number of wires, called lines, and reversible gates that operate on the lines. The lines carry binary values, 0 or 1, which are placed on the lines. The input terminals are on the left side and the output terminals are on the right. All but one of the lines pass through the gate unmodified and are called control lines. The remaining line, called the target line is XORed by the gate with the conjunction of the values of the control lines [9] [10].

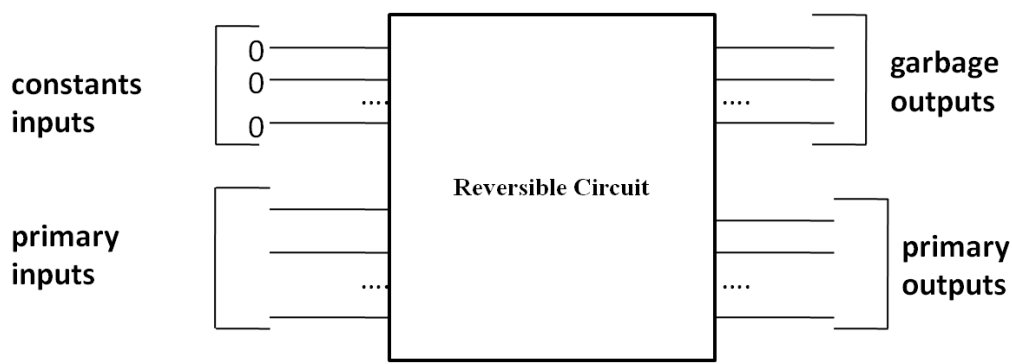

**Fig. 1.2. Reversible Circuit**

### 1.2.3 Application of Reversible Computing

#### 1.2.3.1 Considerations of power consumption

Reversible Computation [7] is a new alternative for Low Power design solutions. So, it is a new area for research. In 1949, John Von Neumann estimated the minimum possible energy dissipation per bit as $k_B T \ln 2$ where $k_B = 1.38065 * 10^{-23}$ J/K is the Boltzmann constant and T is the temperature of environment [Von Neumann 1966]. Subsequently, Landauer [1961] pointed out that the irreversible erasure of a bit of information consumes power and dissipates heat [1]. While reversible designs avoid this aspect of power dissipation, most power consumed by modern circuits is unrelated to computation but is due to clock networks, power and ground networks, wires, repeaters, and memory. There is relation between power consumption and time is shown in Fig. 1.3

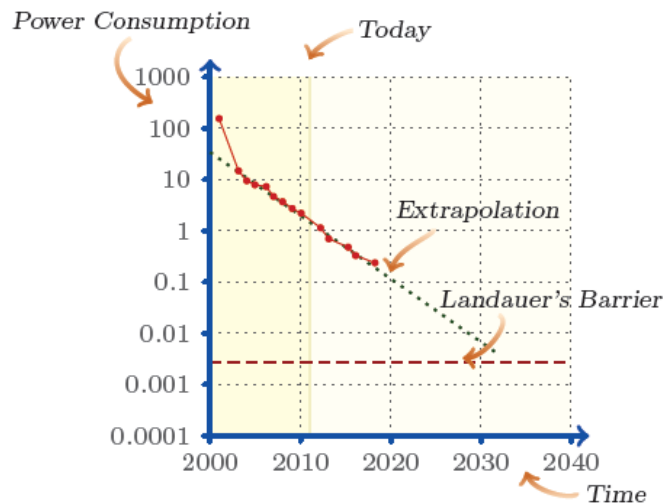

**Fig. 1.3. Relation Between Power Consumption and Time**

### 1.2.3.2 Signal processing, cryptography, and computer graphics

These applications also require reversible transforms, where all of the information encoded in the input must be preserved in the output. A common example is swapping two values a and b without intermediate storage by using bitwise XOR operations a = a⊕b, b = a⊕b, a = a⊕b. In particular, the performance of cryptographic algorithms DES, Twofish and Serpent, as well as string reversals and matrix transpositions, can be considerably improved by the addition of bit-permutation instructions [Shi and Lee 2000; Hilewitz and Lee 2008]. Reversible computations in these applications are usually short and hand-optimized [11].

### 1.2.3.3 Program inversion and reversible debugging

These allow reconstructing sequences of decisions that lead to a particular outcome. Automatic program inversion [Gluke and Kawabe 2005] and reversible programming languages [Yokoyama et al. 2008; De Vos 2010b] allow reversible execution. Reversible debugging [Visan et al. 2009] supports reverse expression watch-pointing to provide further examination of a problematic event [11].

### 1.2.3.4 Networks on chip

Networks on chip with mesh-based and hypercubic topologies [Dally and Towles 2003] perform permutation routing among nodes when each node can both send and receive messages [11].

### 1.2.3.5 Nano- and photonic circuits [Politi et al. 2009; Gao et al. 2010]

These circuits are made up of devices without gain, and they cannot freely duplicate bits because that requires energy. They also tend to recycle available bits to conserve energy. Generally, building nano-size switching devices with gain is difficult because this requires an energy distribution network. Therefore, reversibility is fundamentally important to nanoscale computing, although specific constraints may vary for different technologies [11].

### 1.2.3.6 Quantum computation [Nielsen and Chuang 2000]

Reversible computation is also necessary in application of quantum computation. Quantum algorithms have been designed to solve several problems in polynomial time [Bacon and van Dam 2010; Childs and van Dam 2010], where best known conventional algorithms take more than polynomial time. A key example is number-factoring, which is relevant to cryptography. Research on reversible logic synthesis has attracted much attention after the discovery of powerful quantum algorithms in the mid 1990s [Nielsen and Chuang 2000] [12].

## 1.3 RPGA Structure

In reversible circuit lab label implementation is possible on devices like Programmable logic devices (PLDs)/ Field programmable gate array (FPGA). In PLDs there is a predefined structure available in the form of AND/OR plane. Using some switches, circuit can be realized through fusing binary description of the circuit with the help of some tools. Whereas in FPGA the circuit can be implemented using configurable logic blocks. Hence any circuit which can be accomodated in these devices can be prepared in binary (fusion ready) form using some tools and can be configured in FPGA that FPGA start working like target circuit. Such circuit can be changed/errors can be corrected and results can be seen on this circuit.

In reversible circuit implemented technology is not yet matured. Hence such structures are not available but reversible platform has been proposed consisting of 2 planes. Former plane contains circuit inputs and later plane consists Feynman/Copying gates [13].

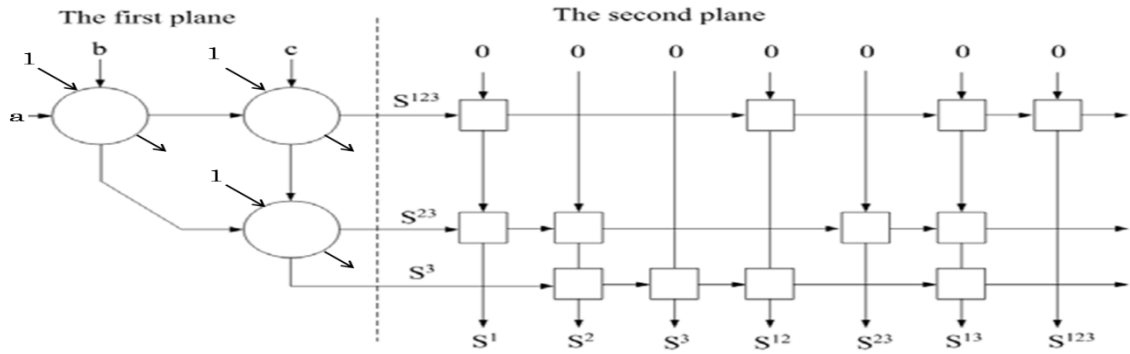

**Fig. 1.4. RPGA structure with Constant/garbage inputs/outputs**

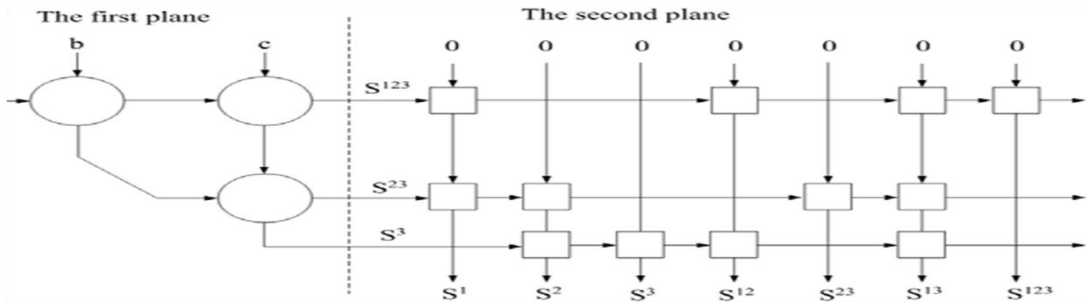

**Fig. 1.5. RPGA structure without Constant/garbage inputs/outputs**

Such structure can implement only symmetric functions. A symmetric function [14] can be defined as that the same number of ones in the input side of the truth table the output will also be same.

**Table 1.2 (a) Symmetric function (b) Asymmetric function**

| a | b | c | out1 | out2 |
|---|---|---|------|------|
| 0 | 0 | 0 | 0 | 0 |
| 0 | 0 | 1 | 1 | 0 |
| 0 | 1 | 0 | 1 | 0 |
| 0 | 1 | 1 | 0 | 1 |
| 1 | 0 | 0 | 1 | 0 |
| 1 | 0 | 1 | 1 | 1 |
| 1 | 1 | 0 | 0 | 1 |
| 1 | 1 | 1 | 1 | 1 |

| a | b | c | out1 | out2 |
|---|---|---|------|------|
| 0 | 0 | 0 | 0 | 0 |
| 0 | 0 | 1 | 1 | 0 |
| 0 | 1 | 0 | 1 | 0 |
| 0 | 1 | 1 | 0 | 1 |
| 1 | 0 | 0 | 0 | 0 |
| 1 | 0 | 1 | 1 | 0 |
| 1 | 1 | 0 | 0 | 1 |
| 1 | 1 | 1 | 0 | 0 |

In section (a), at row 2 input a = 0 b = 0 and c = 1 have output out1 = 1 and at row 3 input a = 0, b = 1 and c = 0 have output out1 = 1 and at row 5 input a =1, b = 0 and c = 0 have output

out1 = 1. Now at row 8 input a = 1, b = 1, c = 1 have output out1 = 1. So, it gives symmetric function at S1,3 for out1 in truth table. Now, at row 4 input a = 0, b = 1, c = 1 have output

out2 = 1, at row 6 input a = 1, b = 0, c = 1 have output out2 = 1, at row 7 input a = 1, b = 1, c = 0 have output out2 = 1 and at row 8 input a = 1, b = 1, c = 1 have output out2 = 1. So, it gives symmetric function at S2,3 for out2 in truth table.

In section (b), at row In section (a), at row 2 input a = 0 b = 0 and c = 1 have output out1 = 1 and at row 3 input a = 0, b = 1 and c = 0 have output out1 = 1 and at row 5 input a =1, b = 0 and c = 0 have output out1 = 0. It gives another output. So, it shows asymmetric function for out1 in truth table. Now, at row 4 input a = 0, b = 1, c = 1 have output out2 = 1, at row 6 input a = 1, b = 0, c = 1 have output out2 = 0, So, it gives asymmetric function for out2 in truth table.

## 1.4 Motivation

Reversible computing is an emerging research area and is a promising technology for the next generation of computers [15]. The primary motivation towards reversible computing is based on the fact that Reversible computing is new alternative for Low power design solution, improved Reversibility, High Packaging density [2]. Traditional computing is near its upper limit. Researcher have shown that reversible computing will also lead to improvement in energy efficiency. Energy efficiency will fundamentally affect the speed of circuits such as nano circuits and therefore the speed of most computing applications. To increase the portability of devices again reversible computing is required. It will let circuit element sizes to reduce to atomic size limits and hence devices will become more portable. The other reason to work on this area is that irreversible computing have many implementation platform like PLD, FPGA, Systolic array and ASIC etc. First two platforms enable circuit implementation in laboraratory instead of foundry/manufacturing place. But in reversible computing adequate tools are yet to develop. Simulation is still under development. So, we need reversible alternative for circuit implementation & testing in lab that may work similar to PLD/FPGA/SYSTOLIC array for reversible circuits.

## 1.5 Objectives

Many synthesis and simulation tools are available for Irreversible computing for platform like PLD, FPGA and SYSTOLIC array etc. But in reversible technology adequate simulation tools are not yet available. The testing of reversible circuits is also a critical aspect in this area, yet this aspect is still underdeveloped. As target implementation technology is not yet matured, the simulation could be a good alternative.

Keeping this in mind we undertook the project and plan to develop a simulation tool with following objectives :

• Entire Project is to be implemented in Java Platform.

• To develop simulation tool using RPGA structure that can simulate any Reversible Circuit.

• Symmetry Analyzer for any reversible circuit.

• The tool can generate RPGA structure of any given input.

• Generate response for any symmetric circuit.

• Can run in step by step mode.

## 1.6 Problem to be addressed in our project

In Dissertation project, we have addressed some problems :

1. Design Entry of a Reversible Circuit.

2. Symmetry Analyzer.

3. Identify a suitable RPGA structure for a given reversible circuit.

4. For a given reversible circuit and given RPGA, simulate the circuit and verify result.

5. Manually create RPGA structure and step by step view the output.

6. Storing and Retrieving in data structure.

## 1.7 Organization of Dissertation

The rest of the dissertation is organized as follows : Chapter 2 provides Literature survey that is useful for getting idea to work on this area. Literature survey describes, reversible logic, reversible circuits, basic reversible logic gates, basic idea of reversible programmable gate array(RPGA), RPGA structure gates, systolic structure, programmable logic array(PLA), and symmetric function. Chapter 3 Design And Development Of RPGA Simulator describes design and development plan of RPGA simulator. The chapter describes the CAD flow, design entry of reversible circuit, Truth table generator, Symmetry analyzer, RPGA structure entry and RPGA simulator and output. This chapter also describes design decision and methodology with algorithms used to implement this work. Chapter 4 Usage And Results Presented GUI interface for design entry of a reversible circuit, GUI interface for Truth Table Generator, GUI interface for symmetry analyzer, and GUI interface for RPGA intial mode, configuration mode and user mode screens. This chapter also describes implementation results. Chapter 5 describes Conclusion And Future Scope of dissertation project. In this chapter we focus on summary of dissertation work and also focus on future directions.

# LITERATURE SURVEY

In this chapter we describe literature survey, that includes background and previous work related to this area. In irreversible computing PLA and FPGA were implementing the circuit in lab. These were having the capabilities to realize/reconfigure hardware to behave like a target circuit. A systolic structure was invented to process the data words from external memory in a rhythmic fashion and achieve a good amount of parallelism with a minimum communication overhead. In order to bring the similar capability with in reversible circuit in RPGA structure has been proposed by Perkowski et. al. group floated an idea of a RPGA based on regular structure to realize binary symmetric functions in reversible logic. We organize our survey in the following categories.

• Reversible Logic Gates.

• Reversible Circuit Design Issues.

• Programmable Logic Array (PLA) and Field Programmable Gate Array (FPGA).

• Systolic Array Structures and Reversible Programmable Gate Array (RPGA).

• Survey Extraction.

## 2.1 Reversible Logic Gates

In this section we, describe reversible logic gates that design a reversible circuits. There are different type of gates in reversible computing such as basic reversible logic gates, complex reversible logic gates, fault tolerant reversible logic gates, conservative reversible logic gates [16] [17] [18].

### 2.1.1 Basic Reversible Logic Gates

In reversible computing there are basic reversible logic gates for develop a reversible circuits. These are as follows.

### 2.1.1.1  Not Gate

Not gate is a 1 * 1 reversible gate is shown in Fig. 2.1. The input is A and the output is P = A'. This is one example of one-bit inverter, with its output being the inverse of its input. Thus, the inverter's input can be regenerated by inverting its output.

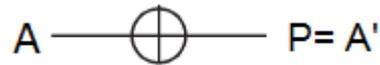

**Fig. 2.1. Reversible Not Gate**

### 2.1.1.2  Feynman Gate (CNOT)

Feynman gate is a 2*2 one-through reversible gate is shown in Fig. 2.2. It is called 2*2 gate because it has 2 inputs and 2 outputs. One through gate means that one input variable is also the output. Here, the input vector is I (A, B) and the output vector is O (P, Q) P = A and Q = A ⊕ B.

Forward Computation:

$$P = A;$$

$$\text{If } A = 0 \text{ then } Q = B$$

$$\text{Else } Q = B'$$

Backward Computation:

$$A = P;$$

$$\text{If } P = 0 \text{ then } B = Q$$

$$\text{Else } B = Q'$$

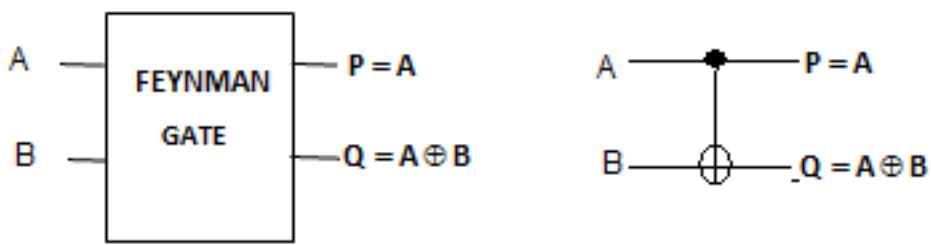

**Fig. 2.2. Reversible Feynman (CNOT) Gate**

**Table 2.1. Truth Table of Reversible Feynman Gate**

| A | B | P | Q |
|---|---|---|---|
| 0 | 0 | 0 | 0 |
| 0 | 1 | 0 | 1 |
| 1 | 0 | 1 | 1 |
| 1 | 1 | 1 | 0 |

### 2.1.1.3 Toffoli Gate

Toffoli Gate ( TG) [17] is a 3*3 two-through reversible gate as shown in Fig. 2.3. In Toffoli Gate, the outputs P and Q are directly generated from inputs A and B respectively by hardwiring. This gate is also suitable for both the forward as well as backward computation [19] [20] [21] .

Forward Computation:

$$P = A; Q = B;$$

If A AND B =0 then R = C

Else R = C'

Backward Computation:

$$A=P; B=Q;$$

If P AND Q=0 then C=R

Else C=R'

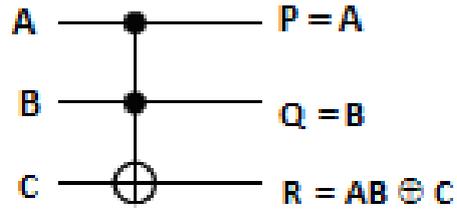

**Fig. 2.3. Reversible Toffoli Gate**

**Table 2.2. Truth Table of Reversible Toffoli Gate**

| A | B | C | P | Q | R |
|---|---|---|---|---|---|
| 0 | 0 | 0 | 0 | 0 | 0 |
| 0 | 0 | 1 | 0 | 0 | 1 |
| 0 | 1 | 0 | 0 | 1 | 0 |
| 0 | 1 | 1 | 0 | 1 | 1 |
| 1 | 0 | 0 | 1 | 0 | 0 |
| 1 | 0 | 1 | 1 | 0 | 1 |
| 1 | 1 | 0 | 1 | 1 | 1 |
| 1 | 1 | 1 | 1 | 1 | 0 |

### 2.1.1.4 Fredkin Gate

Fredkin gate [17] is a ( 3*3) reversible gate originally introduced by Petri as shown in Fig. 2.4. It is called 3*3 gate because it has three inputs and three outputs. The input vector is I(A,B,C) and the output vector is O(P,Q,R) as follows [22] :

Forward Computation:

$$P = A$$

If A=0 then Q=B and R=C

Else Q = C and R=B

Backward Computation:

A=P

If P=0 then B=Q and C=R

Else C=Q and B=R

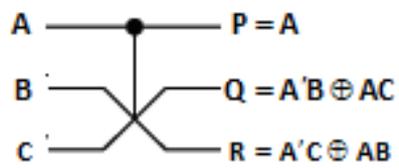

**Fig. 2.4 : Reversible Fredkin Gate**

**Table 2.3. Truth Table of Reversible Fredkin Gate**

| A | B | C | P | Q | R |
|---|---|---|---|---|---|
| 0 | 0 | 0 | 0 | 0 | 0 |
| 0 | 0 | 1 | 0 | 0 | 1 |
| 0 | 1 | 0 | 0 | 1 | 0 |
| 0 | 1 | 1 | 0 | 1 | 1 |
| 1 | 0 | 0 | 1 | 0 | 0 |
| 1 | 0 | 1 | 1 | 1 | 0 |
| 1 | 1 | 0 | 1 | 0 | 1 |
| 1 | 1 | 1 | 1 | 1 | 1 |

## 2.1.1.5 Peres Gate

Peres gate is a ( 3*3) reversible gate as shown in Fig. 2.5. It is called 3*3 gate because it has three inputs and three outputs. The input vector is I(A,B,C) and the output vector is O(P,Q,R).

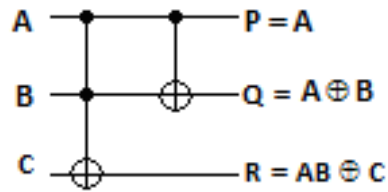

Fig. 2.5. Reversible Peres Gate

Table 2.4. Truth Table of Reversible Peres Gate

| A | B | C | P | Q | R |
|---|---|---|---|---|---|
| 0 | 0 | 0 | 0 | 0 | 0 |
| 0 | 0 | 1 | 0 | 0 | 1 |
| 0 | 1 | 0 | 0 | 1 | 0 |
| 0 | 1 | 1 | 0 | 1 | 1 |
| 1 | 0 | 0 | 1 | 1 | 0 |
| 1 | 0 | 1 | 1 | 1 | 1 |
| 1 | 1 | 0 | 1 | 0 | 1 |
| 1 | 1 | 1 | 1 | 0 | 0 |

## 2.1.2 Complex Reversible Logic Gates

Besides basic reversible logic gates, Reversible computing also contains some complex gates, these are as follows [11] :

### 2.1.2.1 Multiple Control Toffoli Gate

$C^mNOT(x_1; x_2, ......, x_{m+1})$ passes the first m lines, control lines, unchanged. This gate flips the (m+1)-th line, target line, if and only if each positive(negative) control line carries the 1(0) value. For m = 0,1,2 the gates are named NOT(N), CNOT(C), and Toffoli(T), respectively. These three gates compose the universal NCT library.

### 2.1.2.2 Multiple Control Fredkin Gate

Fred($x_1; x_2, ......, x_{m+2}$) has two target lines $x_{m+1}; x_{m+2}$ and m control lines $x_1; x_2, ......, x_m$. The gate interchanges the values of the targets if the conjunction of all m positive(negative) controls evaluates to 1(0). For m = 0,1 the gates are called SWAP(S) and Fredkin(F), respectively.

### 2.1.2.3 Majority Gate and Unmajority and Add UMA gate

The MAJ gate computes the majority of three bits in place [Cuccaro 2005], and provides the carry bit for addition. Cascading it with an Un-majority and Add (UMA) gate [Cuccaro 2005] forms a full adder.

## 2.1.3 Fault Tolerant Gates

There is also some Fault Tolerant Gates. These gates also called parity preserving gates in reversible computation. A few Fault Tolerant gates have been proposed [23] [24] [25].

### 2.1.3.1 Fredkin Gate(FRG)

Following Fig. 2.6 shows a 3*3 FRG gate. The input vector is I(A, B, C) and the output vector is O(P, Q, R). and the output is P = A, Q = A'B ⊕AC, R = A'C ⊕AB.

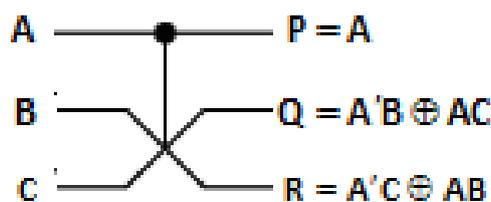

**Fig. 2.6. Fault Tolerant Fredkin Gate**

**Table 2.5. Truth Table of Fault Tolerant Fredkin Gate**

| A | B | C | P | Q | R |
|---|---|---|---|---|---|
| 0 | 0 | 0 | 0 | 0 | 0 |
| 0 | 0 | 1 | 0 | 0 | 1 |
| 0 | 1 | 0 | 0 | 1 | 0 |
| 0 | 1 | 1 | 0 | 1 | 1 |
| 1 | 0 | 0 | 1 | 0 | 0 |
| 1 | 0 | 1 | 1 | 1 | 0 |
| 1 | 1 | 0 | 1 | 0 | 1 |
| 1 | 1 | 1 | 1 | 1 | 1 |

### 2.1.3.2 Double Feynman Gate(F2G)

Following Fig. 2.7 shows a 3*3 F2G gate, there is input vector I(A, B, C) and output vector is O(P, Q, R). the output is P = A, Q = A $\oplus$ B, R= A$\oplus$C

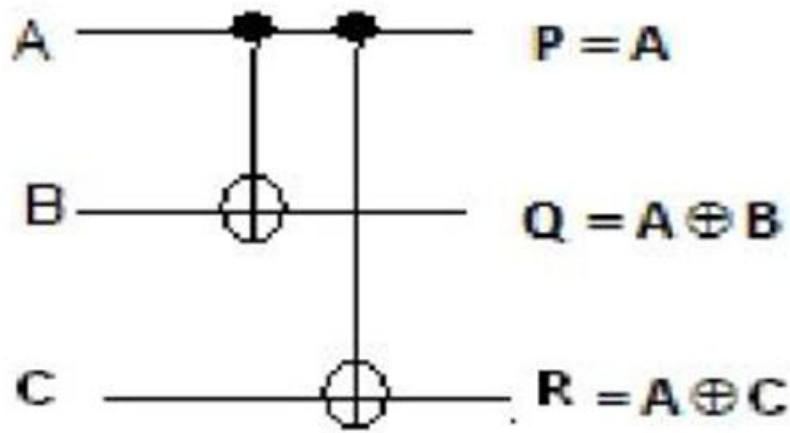

**Fig. 2.7. Fault Tolerant Double Feynman Gate**

**Table 2.6. Truth Table of Fault Tolerant Double Feynman Gate**

| A | B | C | P | Q | R |
|---|---|---|---|---|---|
| 0 | 0 | 0 | 0 | 0 | 0 |
| 0 | 0 | 1 | 0 | 0 | 1 |
| 0 | 1 | 0 | 0 | 1 | 0 |
| 0 | 1 | 1 | 0 | 1 | 1 |
| 1 | 0 | 0 | 1 | 1 | 1 |
| 1 | 0 | 1 | 1 | 1 | 0 |
| 1 | 1 | 0 | 1 | 0 | 1 |
| 1 | 1 | 1 | 1 | 0 | 0 |

### 2.1.3.3 NFT Gate

Following Fig. 2.8 shows a 3*3 NFT gate. There is input vector I(A, B, C ) and output vector O(P, Q, R) [23].

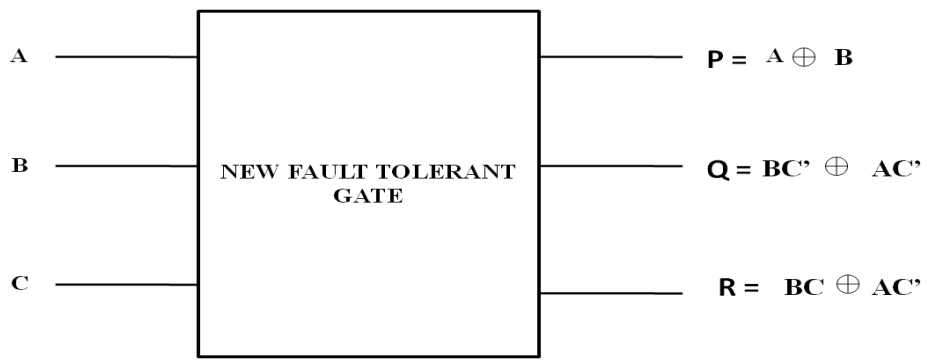

**Fig. 2.8. New Fault Tolerant Gate**

**Table 2.7. Truth Table of New Fault Tolerant Gate**

| A | B | C | P | Q | R |
|---|---|---|---|---|---|
| 0 | 0 | 0 | 0 | 0 | 0 |
| 0 | 0 | 1 | 0 | 1 | 0 |
| 0 | 1 | 0 | 1 | 0 | 0 |
| 0 | 1 | 1 | 1 | 0 | 1 |
| 1 | 0 | 0 | 1 | 1 | 1 |
| 1 | 0 | 1 | 1 | 1 | 0 |
| 1 | 1 | 0 | 0 | 1 | 1 |
| 1 | 1 | 1 | 0 | 0 | 1 |

### 2.1.4 Conservative Reversible Logic Gates

A k*k reversible gate is conservative if the Hamming weight (number of logical ones) of its input equals the Hamming weight if its output. A consequence of a gate's reversibility and conservability is that conservative reversible gates are zero-preserving and one-preserving. A gate is zero-preserving if all zeros input produces all zeros outputs. A gate is one-preserving if all ones input produce all ones outputs. A conservative reversible gate is a gate that is both conservative and reversible. Fredkin Gate is a conservative gate we can easily check it by its truth table [16] [26].

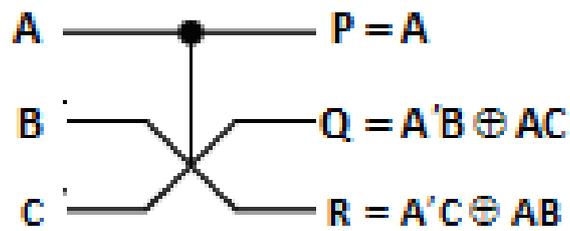

**Fig. 2.9. Conservative Fredkin Gate**

**Table 2.8. Truth Table of Conservative Fredkin Gate**

| A | B | C | P | Q | R |
|---|---|---|---|---|---|
| 0 | 0 | 0 | 0 | 0 | 0 |
| 0 | 0 | 1 | 0 | 0 | 1 |
| 0 | 1 | 0 | 0 | 1 | 0 |
| 0 | 1 | 1 | 0 | 1 | 1 |
| 1 | 0 | 0 | 1 | 0 | 0 |
| 1 | 0 | 1 | 1 | 1 | 0 |
| 1 | 1 | 0 | 1 | 0 | 1 |
| 1 | 1 | 1 | 1 | 1 | 1 |

## 2.2 Reversible Circuit Design Issues

From the point of view of reversible circuit design, there are some design metric for determining the complexity and performance of circuits [9] :

• Number of Reversible Gates(N) : The number of reversible gates used in circuit.

• Number of constant inputs (CI ) : This refers to the number of inputs that are to be maintained constant at either 0 or 1 in order to synthesize the given logical function.

• Number of Garbage outputs (GO) : This refers to the number of unused outputs present in a reversible logic circuit.

• Quantum cost (QC) : This refers to the cost of the circuit in terms of the cost of a primitive gate. It is calculated knowing the number of primitive reversible logic gates (1 * 1 or 2 * 2) required to realize the circuit.

• Gate levels (GL) : This refers to the number of levels in the circuit which are required to realize the given logic functions.

## 2.3 Programmable Logic Array (PLA ) and Field Programmable Gate Array (FPGA)

### 2.3.1 Programmable Logic Array (PLA)

PLA is pre-fabricated building block of many AND/OR gates. This layout allows for a large number of logic functions to be synthesized in the sum of products(and sometimes product of sums). In this figure PLA shows to array of gates one is array of AND gates that takes inputs. After that array of gates it sends the product terms into the array of OR gates. OR gates produce the sum of the previous product terms, that is called sum of product form, and after it outputs generate. Fig. 2.10 shows PLA circuit structure. Here AND gates plane and OR gates plane also shown in the figure. This is called Gate level structure of the PLA [27].

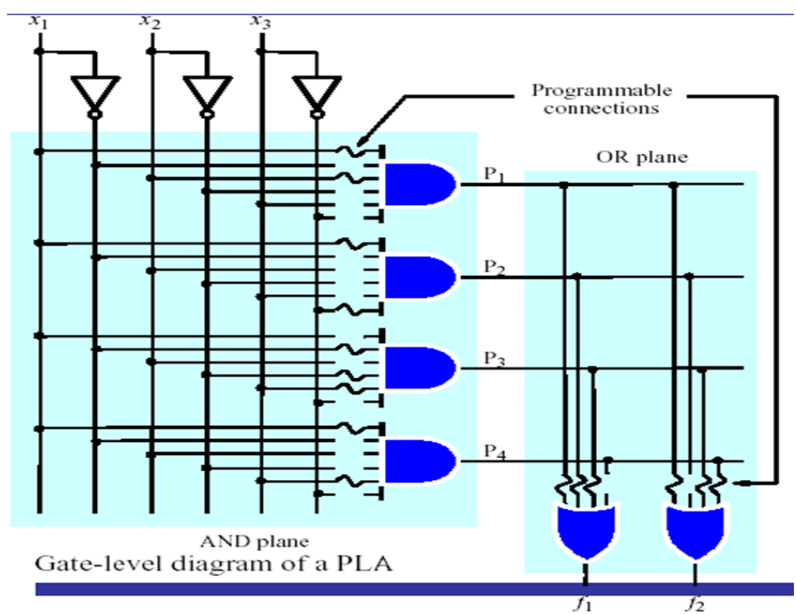

**Fig. 2.10. PLA Circuit Structure**

**2.3.2 Field Programmable Gate Array**

The Field Programmable Gate Array is a device that contains a matrix of reconfigurable gate array logic circuitry. A single FPGA can replace thousands of discrete components by incorporating millions of logic gates in a single integrated circuit (IC) chip. Figure shows the generalize structure of FPGA with Logic block and Input-Output block [27].

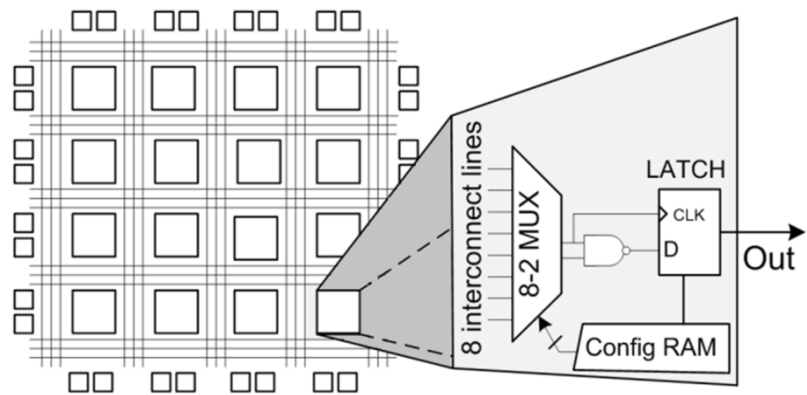

**Fig. 2.11. Fpga Structure**

In the 1980s, FPGAs have grown tremendously in complexity and ubiquitousness. The original Xilinx FPGA contained a mere 64 logic blocks and under 10,000 total transistors. In contrast, variants of the Xilinx Virtex-6 family, released in 2009, contain over 750,000 logic blocks with up to 1,200 I/O pins. Furthermore, most modern FPGAs contain dedicated computation blocks for high-speed digital signal processing (DSP). These range from multipliers and barrel shifters to PCIe interfaces and CPUs. By 2010 the FPGA market is expected to surpass $2.7 billion. These devices are found in innumerable systems from bioinformatics to speech recognition, from ASIC prototyping to radio astronomy. Furthermore, their highly amortized non-recurring engineering costs make FPGAs a cost-effective alternative to custom ASICs.

## 2.4 Systolic Array Structure and Reversible Programmable Gate Array(RPGA)

H.T Kung and C.E. Leiserson introduced the concept of systolic arrays in 1978. The concept of Systolic structure can map high-level computation into hardware structures. Systolic array is a system built with many simple and identical processing elements. These processing elements execute synchronously only basic arithmetic and logic operations. Each processing element is connected with its neighbours only, so interconnections are short, do not cross and form regular structure called mesh. These features allow placing many processing elements in one chip very tightly. Initially, systolic arrays were used as a tool for designing chips in VLSI technology. Systolic structure can result in cost-effective , high- performance special-purpose systems for a wide range of problems [14].

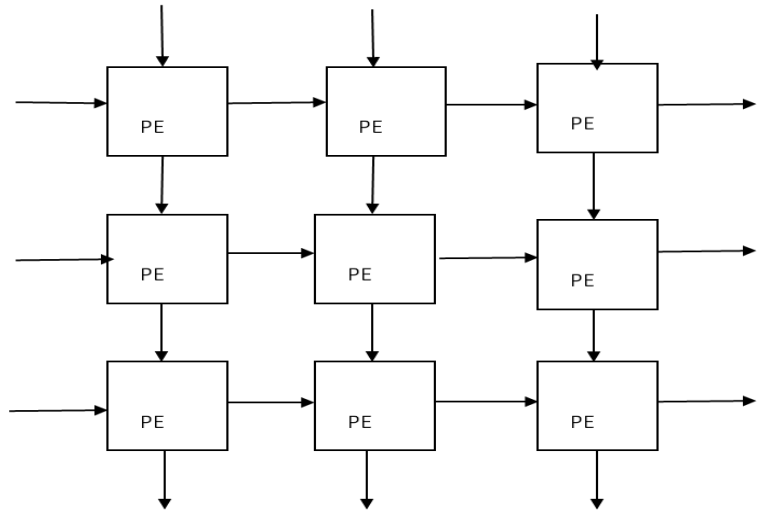

Fig. 2.12. Simple Systolic Array Structure

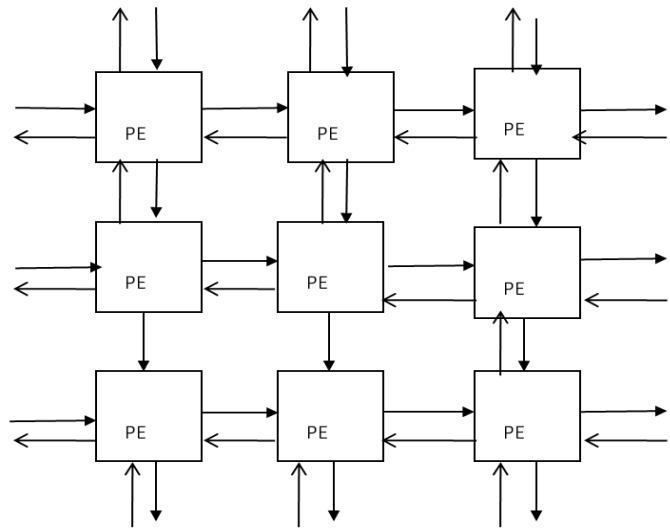

Fig. 2.13. Bidirectional two - dimensional Systolic structure

## 2.4.1 Features of Systolic Array

• Synchrony

• Modularity

• Regularity

• Spatial locality

• Temporal locality

• Pipelinability

• Parallel computing.

• Synchrony : means that the data is rhythmically computed (Timed by a global clock) and passed through the network.

• Modularity : means that the array(Finite/Infinite) consists of modular processing units.

• Regularity : means that the modular processing units are interconnected with homogeneously.

• Spatial Locality : means that the cells has a local communication interconnection.

• Temporal Locality : means that the cells transmits the signals from from one cell to other which require at least one unit time delay.

• Pipelinability : means that the array can achieve a high speed.

### 2.4.2 Reversible Programmable Gate Array(RPGA)

Here, we describe RPGA and systolic array structures.

### 2.4.2.1 RPGA Structure

Portland quantum logic group in [28, 29] floated an idea of a RPGA based on regular structure to realize binary symmetric functions in reversible logic. This structure called a 2*2 Net Structure, allows efficient realization of symmetric functions. The synthesis method to RPGAs allows to realize arbitrary symmetric function in a completely regular structure of reversible gates. A technology independent circuit realization technique was presented by Perkowski et. al. [14, 28]. Hence this technique can be used in association with any known or future target technology for implementation. They believed, however, that regularity of this networks will constitute an additional asset for the forthcoming technoogies, especially nano-technologies.

Fig. 2.14 presents a regular structure based on the three regular planes. The structure of the first plane is planer, regular Symmetric Structure. It realizes all positive unate Symmetric Functions of its input variables, and after the first level, EXOR - levels are shown in this figure. These EXOR levels are the pair of columns of Feynman gates that realizes Single Index and other arbitrary symmetric function. The functionality of these plane can be compared to that of the tradional PLA structure, that have AND/OR plane, used to realize a Sum-of-Product (SOP,DNF) expression for Irreversible Computing. here, positive polarity unate symmetric functions are created systematically in a regular planar arrangement of MAX/MIN modules [28, 29].

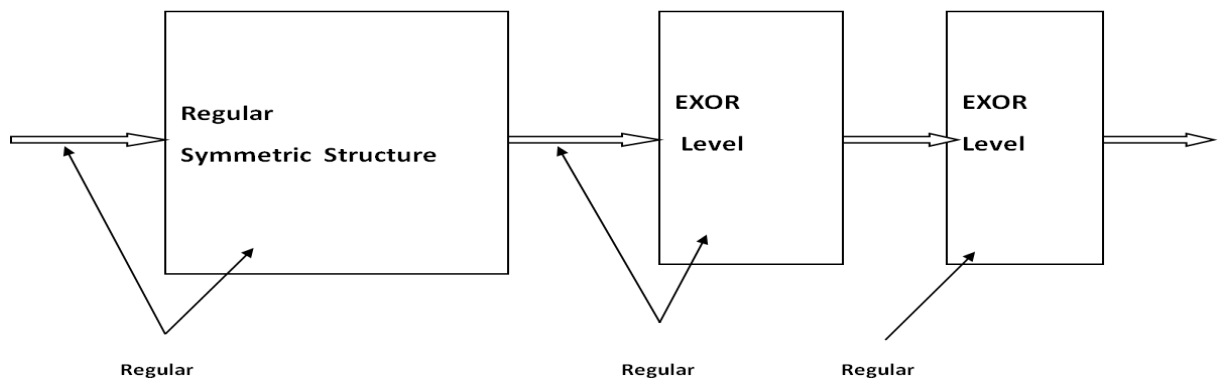

**Fig. 2.14. RPGA block diagram**

Fig. 2.15 illustrates how positive polarity unate symmetric functions are created systematically in a regular planar arrangement of MAX/MIN modules. This is the block diagram of a first plane of a RPGA strucutre [28] that presented by previous author. The first plane is a levelized triangular structure in which the input variables correspond to the columns. This is also called triangular plane. The plane realizes symmetric functions in its outputs, and gives OR/AND (MAX/MIN) [28, 29] combination is shown in below figure. Observe that each output function, from top to bottom, is positively polarity unate and includes the next function. The sets of indices of the adjacent functions differ by one.

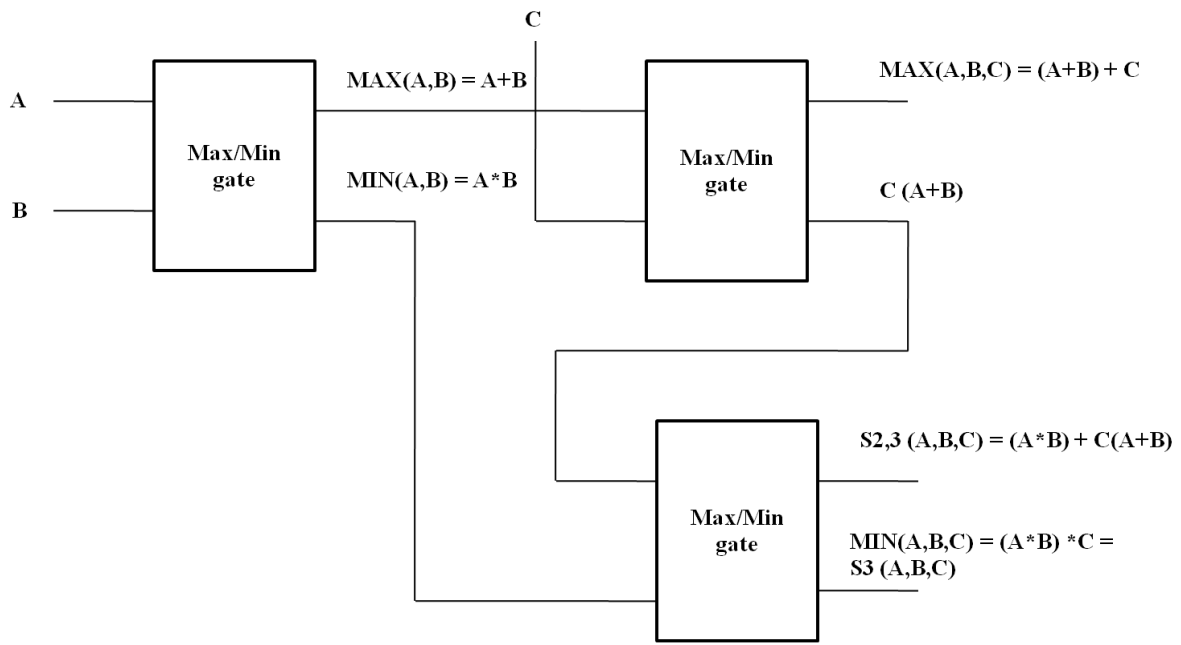

**Fig. 2.15. An Example of Realization of Some Symmetric Functions of Three Input Variables**

Fig. 2.16 shows realization of all single-index symmetric function using only reversible gates as EXORs in the left plane. First plane shows the regular structure from reversible MAX/MIN gates realizing positive polarity unate symmetric functions. Second plane shows that the positive unate symmetric functions generated on the outputs of the triangular plane have a nice property: the EXOR of the neighbor functions creates a single-index symmetric function. This is illustrated in Fig. 2.16. However, because the EXOR gate is not reversible, researcher completed it to Feynman gate by repeating one of its inputs to the output. Because this structure is regular, this does not complicate the structure. In this regular structure, previous researcher obtained not only the single index symmetric functions, but also some interval symmetric functions.

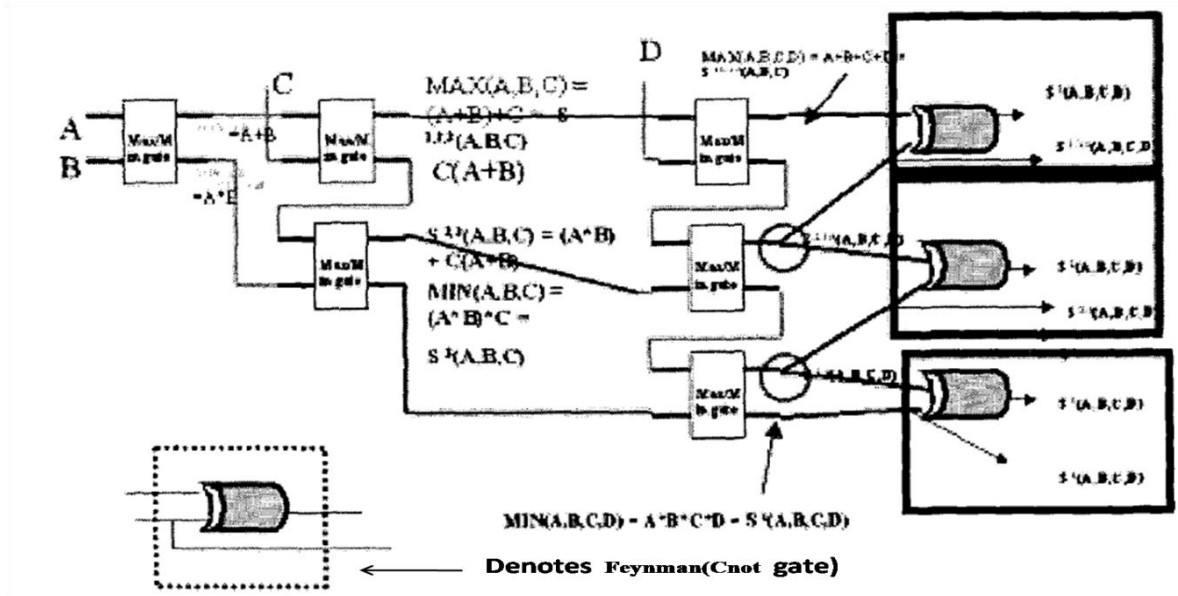

**Fig. 2.16. Realization of all single-index symmetric function using only reversible gates as EXORs**

### 2.4.3 RPGA Structure Gates

To realize RPGA structures following reversible gates are used:

### 2.4.3.1 Picton Gate

Picton Gate in [28,29] published by Perkowski. This is a 4 input and 4 output gate as shown in Fig. 2.17. Previous researcher showed its usefulness to build MAX/MIN-based Sum-of-product realizations in the reversible logic. So, it can be used to build first plane of RPGA structure. The input vector is I(A,B,C,D) and the output vector is O(P,Q,R,S). The proposed designs are often quite inefficient when applied to the quantum logic. Output equation is given as follows.

P = A

Q = B

R=C if A<B else R=D

S=D if A<B else S=C

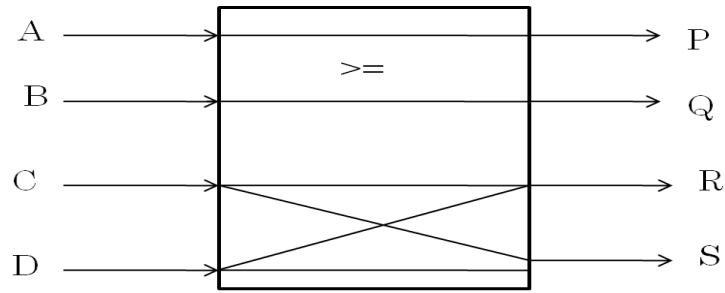

**Fig. 2.17. Reversible Picton Gate**

**Table 2.9. Truth Table of Reversible Picton Gate**

| A | B | C | D | P | Q | R | S |
|---|---|---|---|---|---|---|---|
| 0 | 0 | 0 | o | 0 | 0 | 0 | 0 |
| 0 | 0 | 0 | 1 | 0 | 0 | 1 | 0 |
| 0 | 0 | 1 | 0 | 0 | 0 | 0 | 1 |
| 0 | 0 | 1 | 1 | 0 | 0 | 1 | 1 |
| 0 | 1 | 0 | 0 | 0 | 1 | 0 | 0 |
| 0 | 1 | 0 | 1 | 0 | 1 | 0 | 1 |
| 0 | 1 | 1 | 0 | 0 | 1 | 1 | 0 |
| 0 | 1 | 1 | 1 | 0 | 1 | 1 | 1 |
| 1 | 0 | 0 | 0 | 1 | 0 | 0 | 0 |
| 1 | 0 | 0 | 1 | 1 | 0 | 1 | 0 |
| 1 | 0 | 1 | 0 | 1 | 0 | 0 | 1 |
| 1 | 0 | 1 | 1 | 1 | 0 | 1 | 1 |
| 1 | 1 | 0 | 0 | 1 | 1 | 0 | 0 |
| 1 | 1 | 0 | 1 | 1 | 1 | 1 | 0 |
| 1 | 1 | 1 | 0 | 1 | 1 | 0 | 1 |
| 1 | 1 | 1 | 1 | 1 | 1 | 1 | 1 |

### 2.4.3.2 Picton as a MAX/MIN Gate

Picton gate used to create MAX/MIN gate.in this structure, MAX/MIN Gate can be built using two Picton Gate. Here two constant input 0,1 at the first picton Gate. Output of first gate sends to the second Picton gate, and then 4 outputs are shown in figure. Here two outputs are MAX and MIN and other two are garbage. So, in this structure, it gives two garbage output.

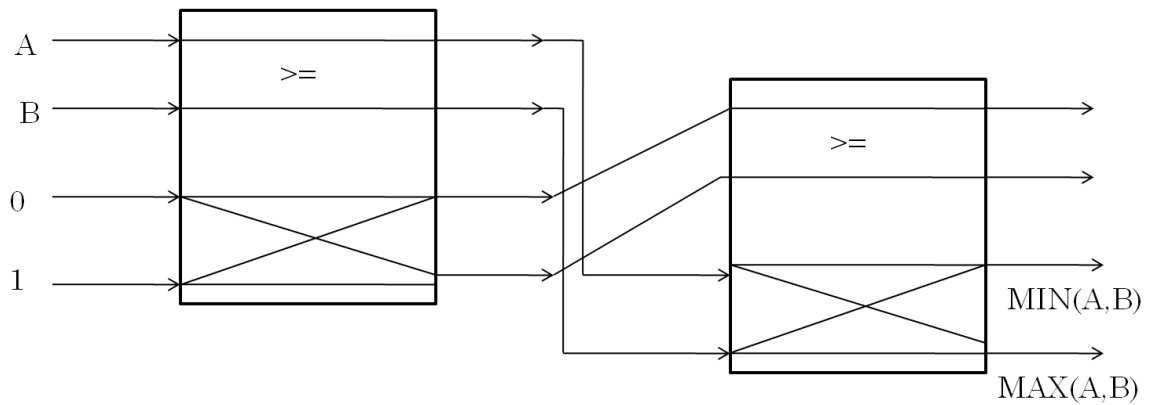

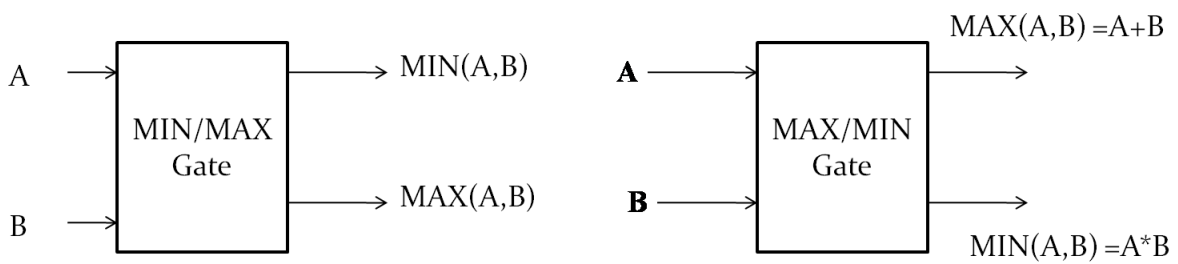

**Fig. 2.18. Picton as MAX/MIN Gate**

**Table 2.10. Truth Table of Picton as MAX/MIN Gate**

| A | B | C | D |
|---|---|---|---|
| 0 | 0 | 0 | 0 |
| 0 | 1 | 1 | 0 |
| 1 | 0 | 1 | 0 |
| 1 | 1 | 1 | 1 |

### 2.4.3.3 Kerntopf Gate

Kerntopf Gate [28,29] is described by equations : P = 1@A@B@C@AB, Q = 1@AB@B@C@BC, R = 1@A@B@AC, where @ denotes EXOR.

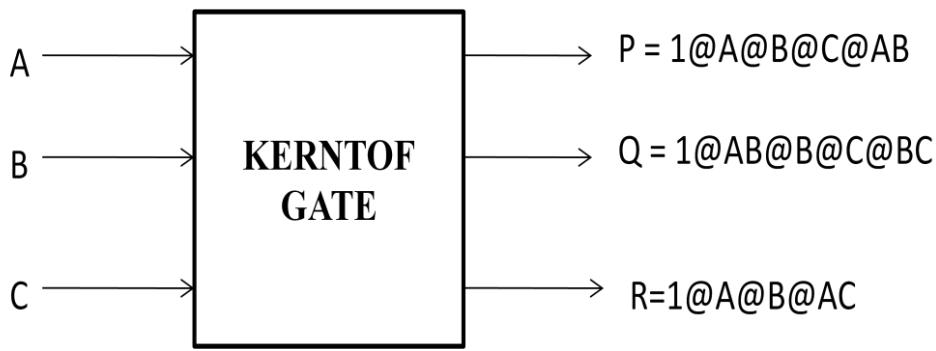

**Fig. 2.19. Reversible Kerntopf Gate**

Table 2.11. Truth Table of Reversible Kerntopf Gate

| A | B | C | P | Q | R |
|---|---|---|---|---|---|
| 0 | 0 | 0 | 1 | 1 | 1 |
| 0 | 0 | 1 | 0 | 0 | 1 |
| 0 | 1 | 0 | 0 | 0 | 0 |
| 0 | 1 | 1 | 1 | 0 | 0 |
| 1 | 0 | 0 | 0 | 1 | 0 |
| 1 | 0 | 1 | 1 | 0 | 1 |
| 1 | 1 | 0 | 0 | 1 | 1 |
| 1 | 1 | 1 | 1 | 1 | 0 |

### 2.4.3.4 Kerntopf as a MAX/MIN Gate

In Kerntopf gate there is 3 inputs A, B and C. So, there is $2^n$ combination of inputs in its truth table. When input C = 1 for all combination of inputs, then P = A+B, Q = A*B, R = !C, so MAX/MIN gate is realized on outputs P and Q with C as the controlling input value. So, for RPGA structure we need MAX/MIN gate in its first plane. Here kerntopf gives MAX/MIN output and R shows garbage bit then it is used in a RPGA structure. So, it gives one Garbage bit, so this gate is better than the Picton Gate that gives two garbage bits. So, kerntopf gate is a better option to create a RPGA structure [28,29].

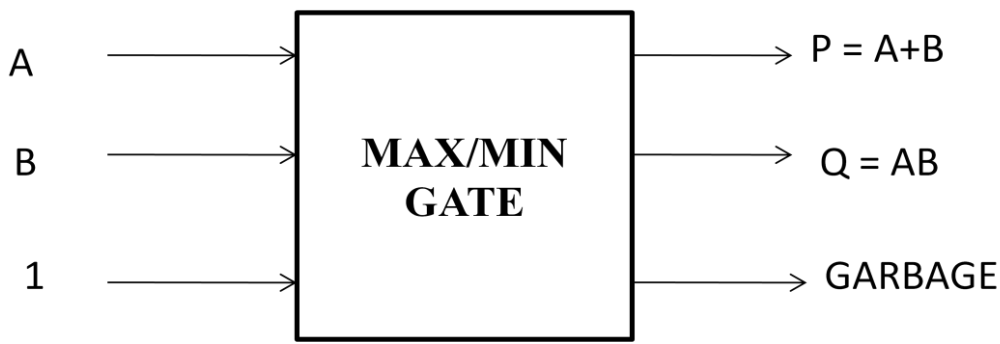

Fig. 2.20. Kerntopf as a MAX/MIN Gate

**Table 2.12. Truth Table of Kerntopf as MAX/MIN Gate**

| A | B | C | P | Q | R |
|---|---|---|---|---|---|
| 0 | 0 | 1 | 0 | 0 | 1 |
| 0 | 0 | 1 | 0 | 0 | 1 |
| 0 | 1 | 1 | 1 | 0 | 0 |
| 0 | 1 | 1 | 1 | 0 | 0 |
| 1 | 0 | 1 | 1 | 0 | 1 |
| 1 | 0 | 1 | 1 | 0 | 1 |
| 1 | 1 | 1 | 1 | 1 | 0 |
| 1 | 1 | 1 | 1 | 1 | 0 |

## 2.5  Survey Extraction

From our survey, it is observed that RPGA structure can implement in lab any symmetric circuit whether (reversible or irreversible) using only reversible structures. As hardware for RPGA is non-existent there is need to develop a simulation of RPGA structures which can interactively implement symmetric function. Of course a symmetry analyzer is also needed before implementation. This project also includes symmetry analyzer and entire RPGA development framework.

# DESIGN AND DEVELOPMENT OF RPGA SIMULATOR

After literature survey, we planned to develop a simulation tool named "RPGA SIMULATOR".

The detailed plan of the tool is described here :

1. Design Entry of a reversible circuit

2. Symmetry Analyzer

3. Identify a suitable RPGA structure for a given circuit

4. For a given circuit and given RPGA simulate the circuit and verify result

5. Manually create RPGA structure and step by step view the output.

The major CAD flow that uses above features in the planned tool is shown in Fig. 3.1.

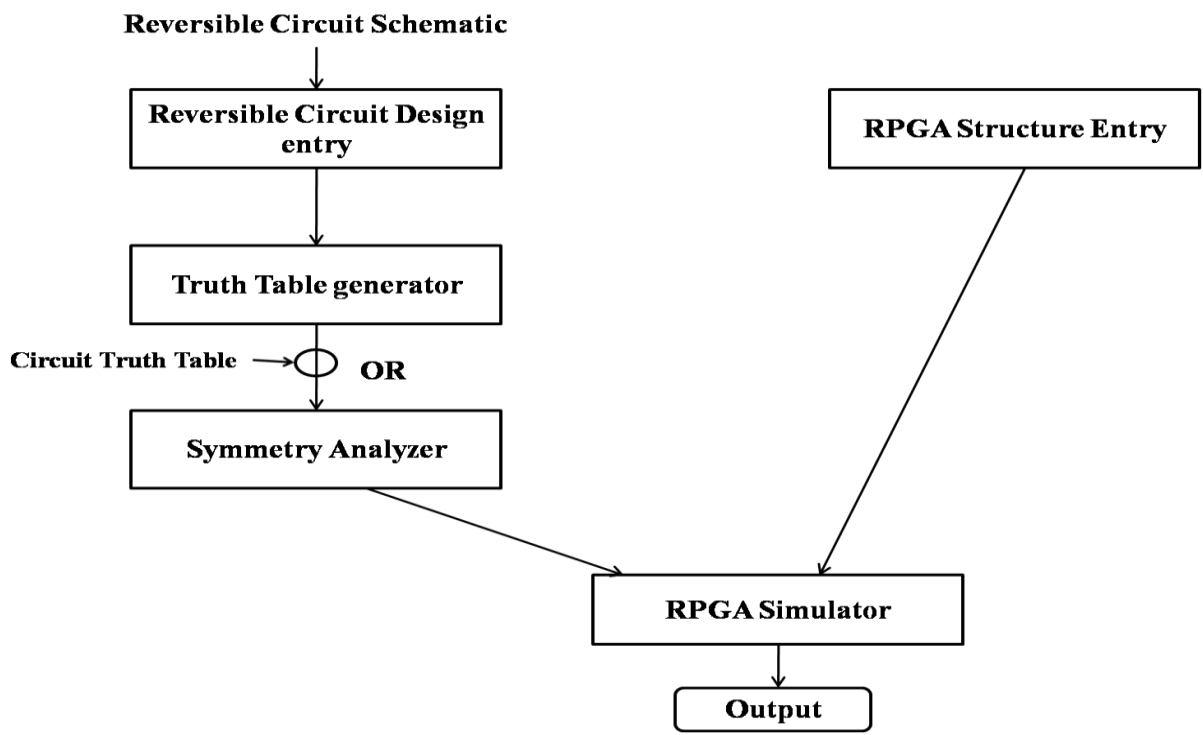

Fig. 3.1. CAD flow for RPGA Simulation

The tool should allow reversible circuit design entry from its schematic followed by truth table generation module. As only truth table is needed for simulation purpose direct entry of truth table is also allowed. By this way truth table of even irreversible circuit can also be used in RPGA simulator.

The truth table then analyzed for its symmetry before implementation in RPGA structure. Each block of above CAD flow shown in Fig. 3.1 will involve a number of steps describe in the following sections.

## 3.1 Design Entry of a reversible circuit

In this section we describe that how to enter design of a reversible circuit in RPGA simulator. The schematic of a design should be readily available. The entire circuit is divided in time slot in such a way so that on any line there should be at most one gate in one slot and the facility should be available to pick the gate from gate library. A good data structure should capture the design for storage and retrieval purpose.

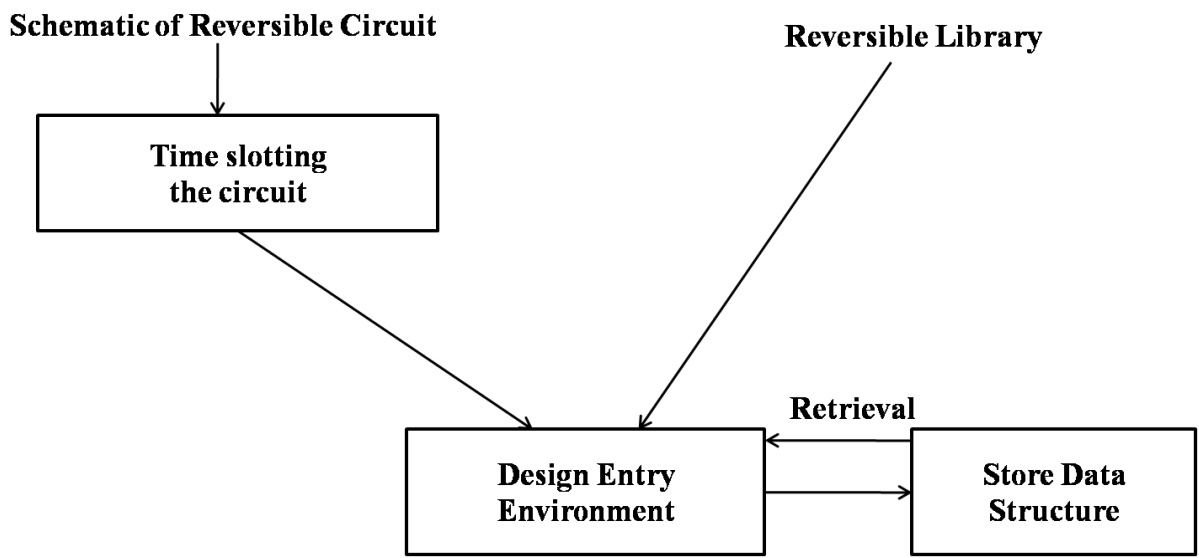

**Fig. 3.2. Design Entry of a Reversible Circuit**

Fig. 3.2 shows the environment for design entry. There are one standard library are available in the literature namely NCT.

The library contains NOT, CNOT and Toffoli reversible logic gates which is commonly used. For normal circuits NCT library is sufficient and easy to used. In our project we have considered NCT library. This gate are described in literature survey in Fig. 2.1, 2.2, 2.3.

## 3.2 Truth Table Generator

Next step of CAD flow in Fig. 3.1 is Truth Table generation. As we know the truth table of reversible circuit is different then the truth table of irreversible circuit. Truth Table of reversible circuit includes constant/garbage inputs/outputs but the truth table of irreversible circuit does not contain these two. The RPGA simulator requires irreversible truth table for implementation of the circuit through reversible structures. Hence the generation of truth table involves two steps namely generation of reversible truth table followed by generation of irreversible truth table as in Fig. 3.3.

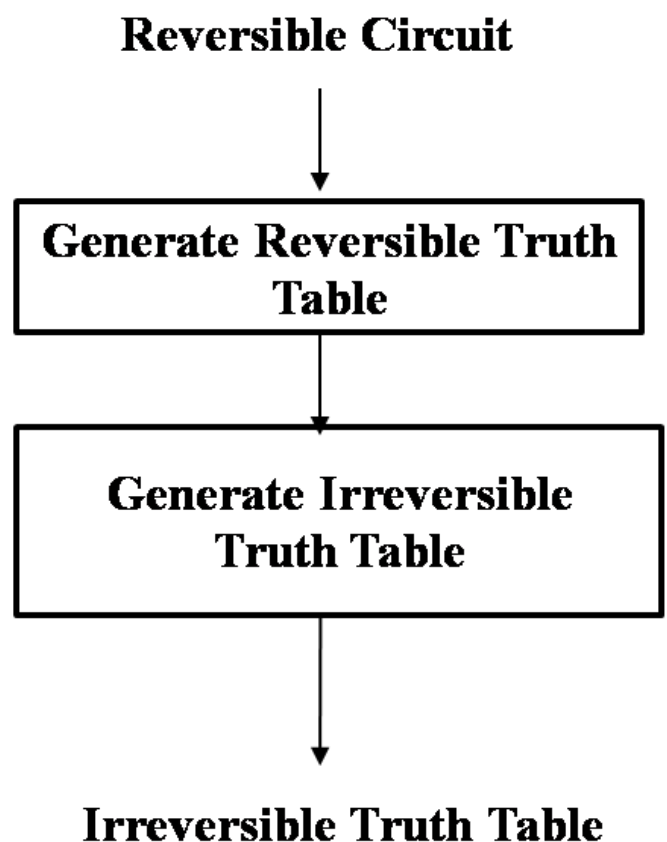

**Fig. 3.3. Generation of Irreversible Truth Table**

In step 1 reversible truth table is generated from the displayed circuit after design entry

The step no. 2 removes columns related to constant inputs/garbages. Fig. 3.4(a) shows reversible circuit and Fig. 3.4(b) shows its reversible truth table and Fig. 3.4(c) shows irreversible truth table after removal of garbage/constant inputs.

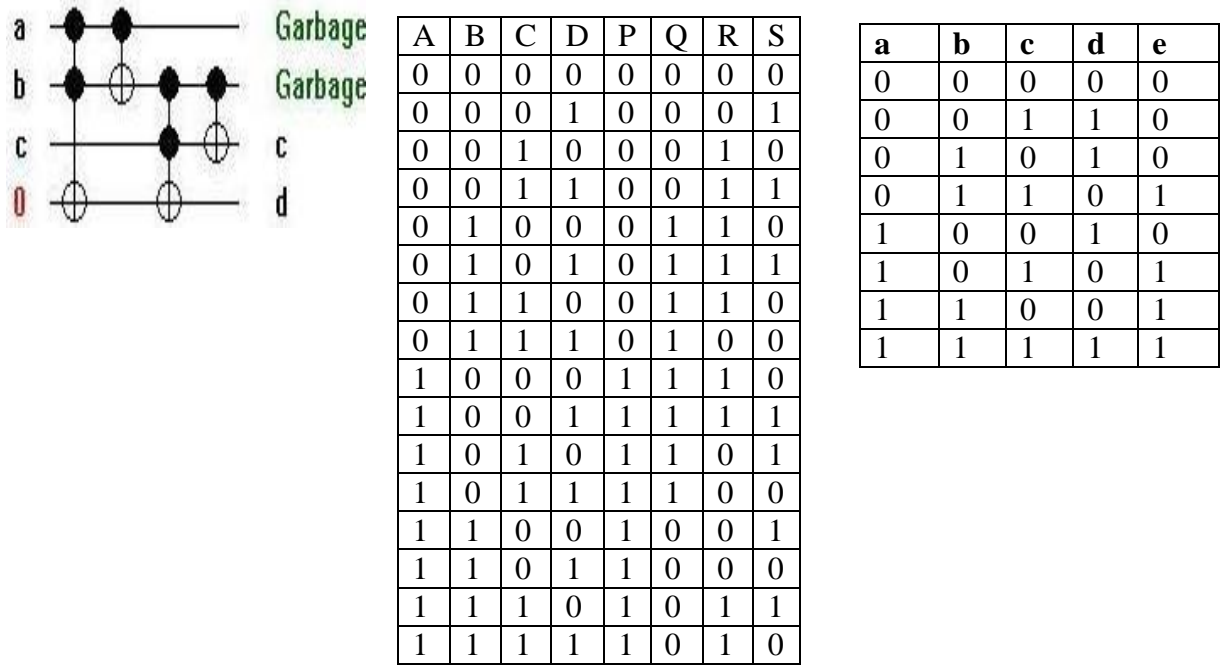

| A | B | C | D | P | Q | R | S |
|---|---|---|---|---|---|---|---|
| 0 | 0 | 0 | 0 | 0 | 0 | 0 | 0 |
| 0 | 0 | 0 | 1 | 0 | 0 | 0 | 1 |
| 0 | 0 | 1 | 0 | 0 | 0 | 1 | 0 |
| 0 | 0 | 1 | 1 | 0 | 0 | 1 | 1 |
| 0 | 1 | 0 | 0 | 0 | 1 | 1 | 0 |
| 0 | 1 | 0 | 1 | 0 | 1 | 1 | 1 |
| 0 | 1 | 1 | 0 | 0 | 1 | 1 | 0 |
| 0 | 1 | 1 | 1 | 0 | 1 | 0 | 0 |
| 1 | 0 | 0 | 0 | 1 | 1 | 1 | 0 |
| 1 | 0 | 0 | 1 | 1 | 1 | 1 | 1 |
| 1 | 0 | 1 | 0 | 1 | 1 | 0 | 1 |
| 1 | 0 | 1 | 1 | 1 | 1 | 0 | 0 |
| 1 | 1 | 0 | 0 | 1 | 0 | 0 | 1 |
| 1 | 1 | 0 | 1 | 1 | 0 | 0 | 0 |
| 1 | 1 | 1 | 0 | 1 | 0 | 1 | 1 |
| 1 | 1 | 1 | 1 | 1 | 0 | 1 | 0 |

| a | b | c | d | e |
|---|---|---|---|---|
| 0 | 0 | 0 | 0 | 0 |
| 0 | 0 | 1 | 1 | 0 |
| 0 | 1 | 0 | 1 | 0 |
| 0 | 1 | 1 | 0 | 1 |
| 1 | 0 | 0 | 1 | 0 |
| 1 | 0 | 1 | 0 | 1 |
| 1 | 1 | 0 | 0 | 1 |
| 1 | 1 | 1 | 1 | 1 |

**Fig. 3.4. (a) reversible circuit (b) reversible truth table (c) irreversible truth table**

Fig. 3.5 shows the conversion of circuit to truth table from design entry. For n inputs there are $2^n$ possible input combinations. As the number of inputs is not a large number $2^n$ combinations can be written in the input portion of truth table. The truth table will contain 2n columns for n- input circuits. Each column will be filled with $2^n$ entries represented in binary.

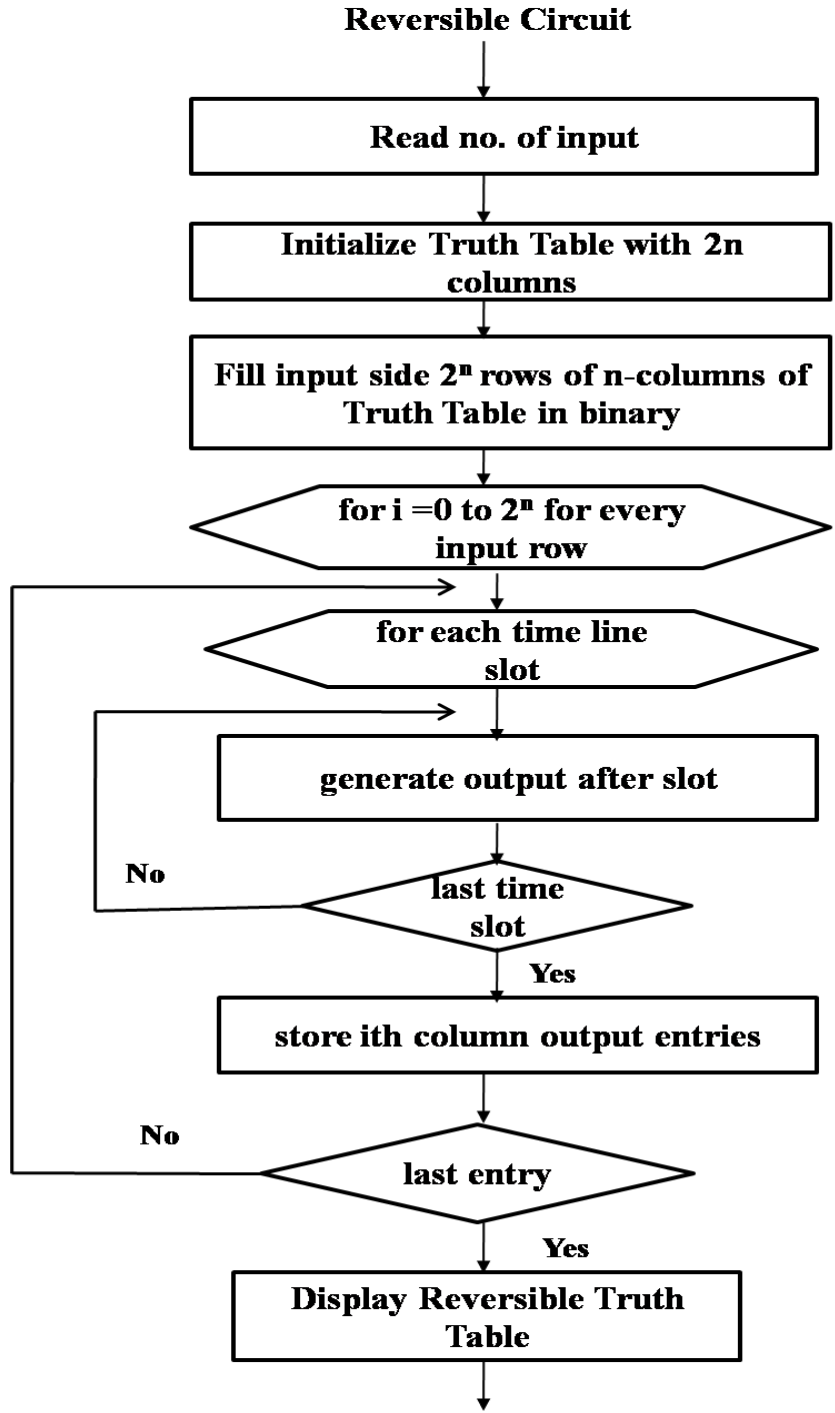

**Fig. 3.5. Conversion of Circuit to Truth Table from Design Entry**

As per our plan tool is also capable of processing irreversible truth table also. Hence the user of the tool will have both the options for truth table generation entry.

**Algorithm 3.1. Algorithm for conversion of circuit to truth table from design entry**

Algo : This algorithm will generate the Truth Table for a reversible circuit

Input : Reversible circuit from design entry.

Output : generated Truth Table.

1. Read number of input

2. initialize Truth table with 2n column.

3. Fill input side $2^{n}$ rows of n- columns of truth table in binary

4. for i = 0 to $2^{n}$ for every input row repeat step 5 to 9

5. for each time line slot repeat step 6 to 7

6. generate output after slot

7. check last time slot

8. Store ith column output entries

9. check last entry

10. Display reversible truth table

11. Stop.

## 3.3 Symmetry Analyzer

As the RPGA structure capable of simulating only symmetric circuits, analysis of the symmetry in truth table of the circuit is necessary. Fig. 3.6 describes the symmetry analysis of irreversible truth table of a circuit.

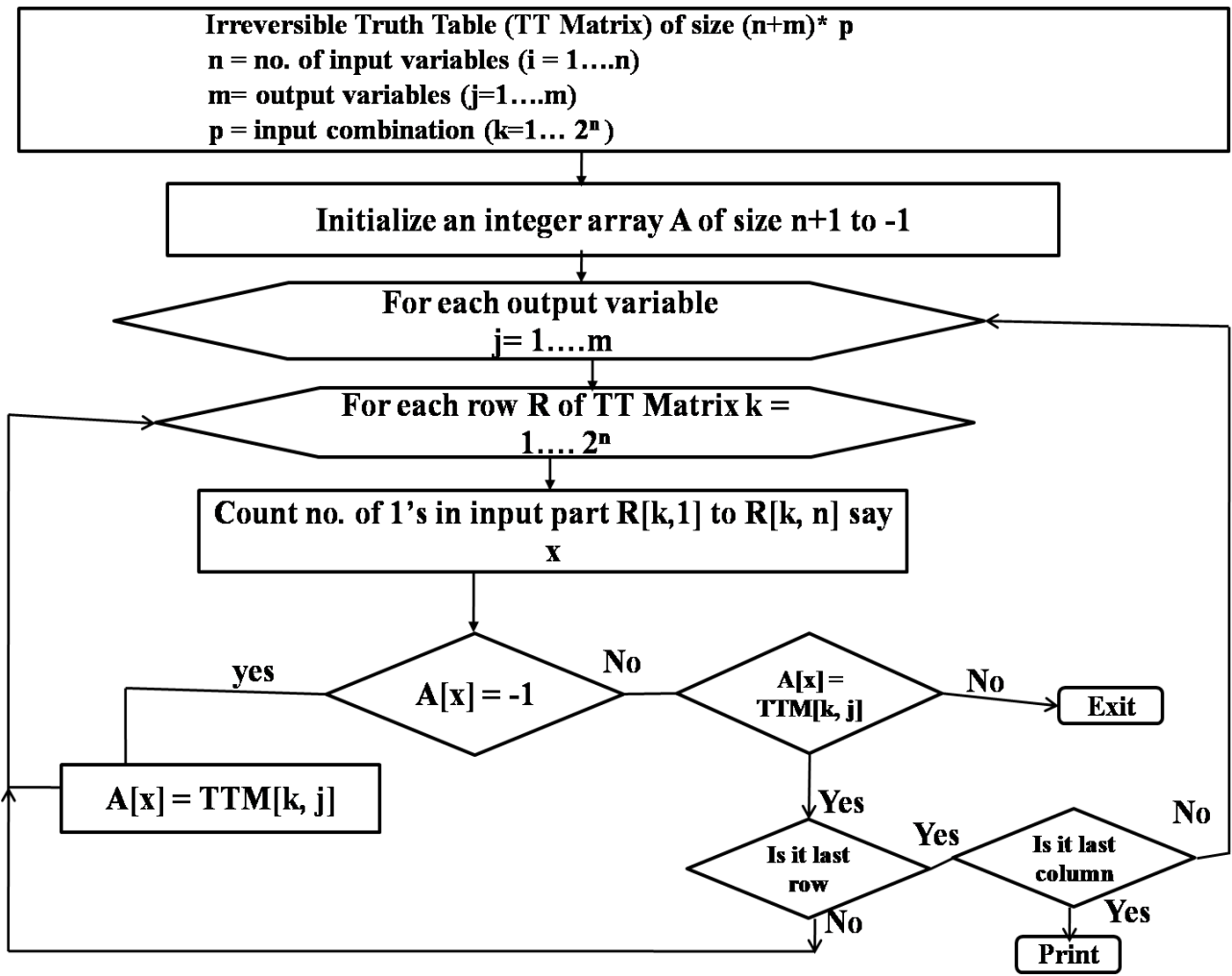

Fig. 3.6. Symmetry Analysis

---

Algorithm 3.2. Algorithm for Symmetry Analysis

---

Algo : This algorithm perform the symmetry analysis of irreversile truth table of circuit

Input : irreversible truth table (TT Matrix) of size (n+m)*p

n = number of input variables (i = 1.......n)

m = output variables (j = 1......m)

p = input combinations (k = 1...... $2^n$

Output : Symmetric results.

1. initialize an integer array A of size n+1 to -1.

2. for each output variable j = 1.....m repeat step 3 to 9.

3. for each row R of TT Matrix k = 1......$2^n$

  repeat step 4 to 8.

4. count number of 1's in input part R[k, 1] to R[k, n ] say x

5. Check A[x] = -1

6. then A[x] = TTM[k, j]

7. Otherwise check A[x] = TTM [k, j]

8. Check is it last row

9. Check is it last column

10. print results

11. Stop.

---

## 3.4 RPGA Structure Entry

There are two planes in RPGA structure [13]. Former plane contains circuit inputs and complex gates like (Picton [13], Kerntopf [13]). The structure was shown in Chapter 1 in Fig. 1.5 however for ready reference Fig. 3.7 it reproduce.

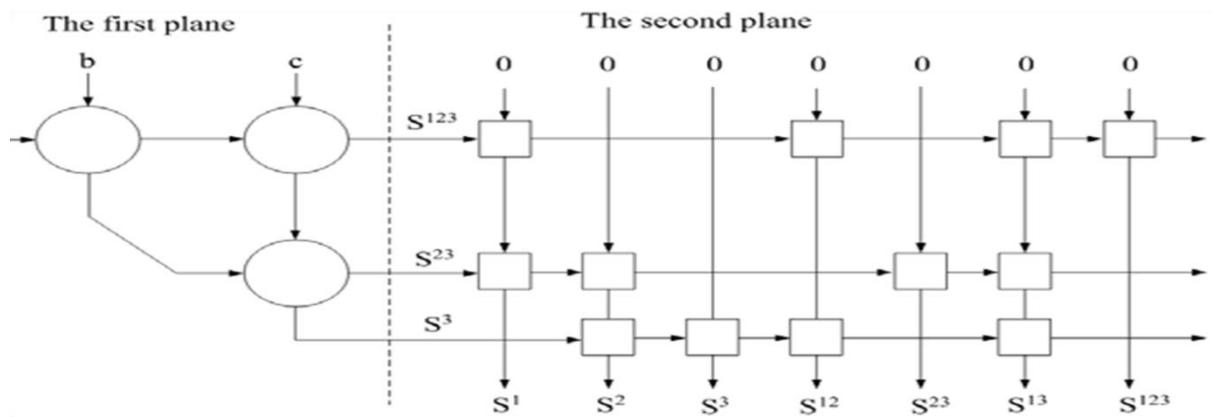

**Fig. 3.7. RPGA structure**

Basically in former plane MAX/MIN out of the two inputs is desired. This MAX/MIN has been achieved in kerntopf gate by making third input c = 1, in picton gate by making third and forth input. However later plane requires copying and Feynman gates. The inputs are fed to plane one and outputs are taken out from later plane. For n - inputs there are $2^{n}$ possible combinations. Input lines are fed at the input of first plane. Hence required n - input lines however output is obtained at one of the output lines. To represent output for every symmetric function $2^n - 1$ (for n = 4, $2^n - 1 = 15$ ) lines in output. The tool is expected to draw entire RPGA structure based on number of input lines. These lines will represent the output corresponding to one function. As multiple outputs are available on this structure, multiple functions can also be implemented.

## 3.5 RPGA Simulator and Output

The simulator program will display RPGA structure assuming MAX/MIN, Copying and Feynman gate. The MAX/MIN Gate is derived from other complex gates but our simulator uses as MAX/MIN gate. The simulator takes as at input a truth table (after symmetry analysis) and highlights the outputs.

### 3.5.1 Configuration and User mode

RPGA simulator works into 2 modes namely

1. Configuration Mode

2. User Mode

The two modes are described below:

## 3.5.1.1 Configuration Mode

RPGA simulator starts in configuration mode all its outputs are inactive Configuration identifies right output pin as per input circuit. Fig. 3.8(a) shows RPGA and Fig. 3.8(b) configured RPGA. Initially input plane, lines and output plane are of black color. When configured the input plane are of pink color, input and output lines are of yellow color, inactive blocks are of light gray color and active blocks are of orange color. In Figure there are two active blocks S1,3 = O1 and S2,3 = O2.

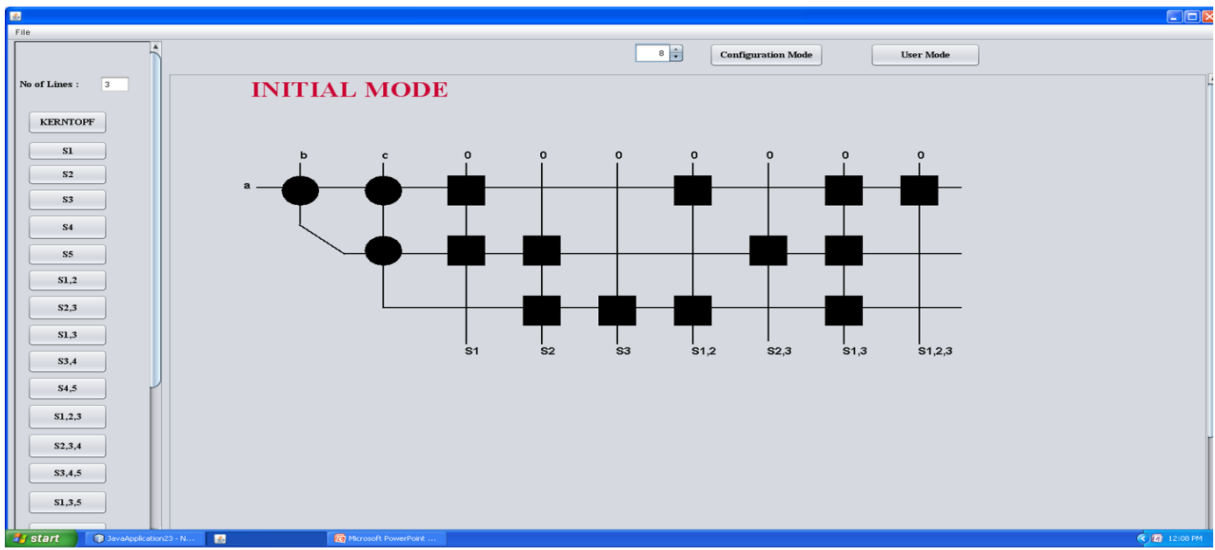

Fig. 3.8(a). RPGA structure

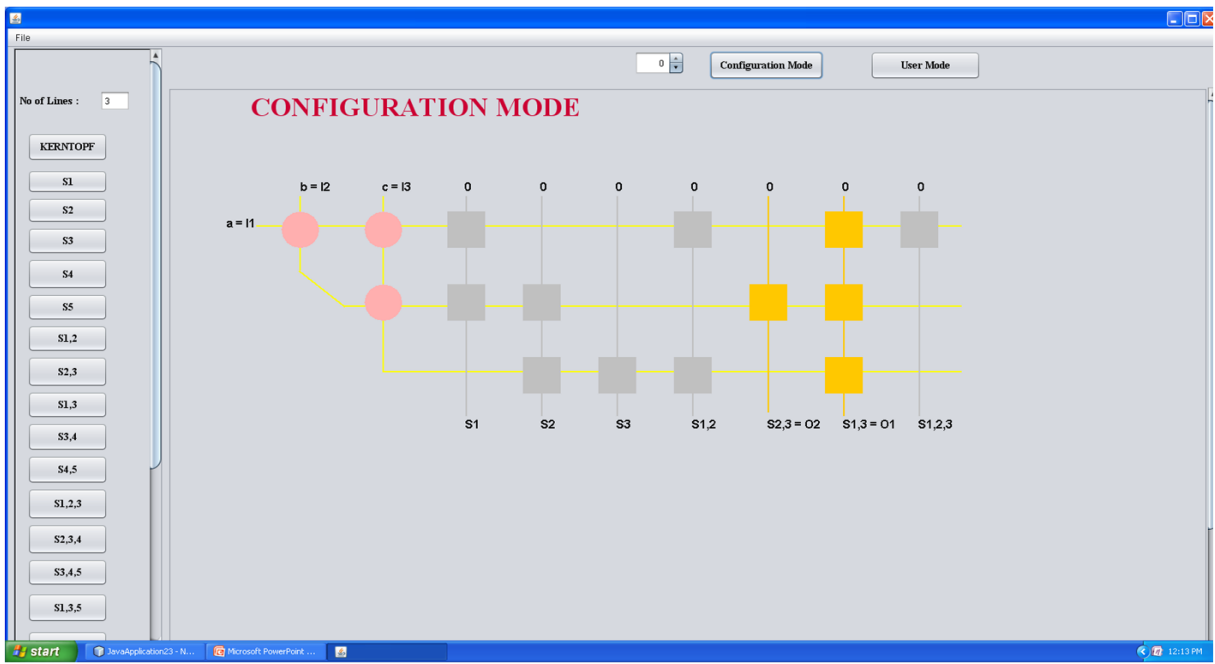

Fig. 3.8(b). Configured RPGA

**3.5.1.2 User Mode**

Once the RPGA is configured for desired multioutput circuit. User can apply desired inputs and get the output. The user can also test the output for all possible inputs through next/previous button. In Fig. 3.9 suppose user apply input I1 = 1, I2 = 0, I3 = 0 then user can get the outputs at S1,3 = O1 and S2,3 = O2. So, for I1 = 1, I2 = 0 and I3 = 0 output line S1,3 shows green(1) color and output line shows red(0) color.

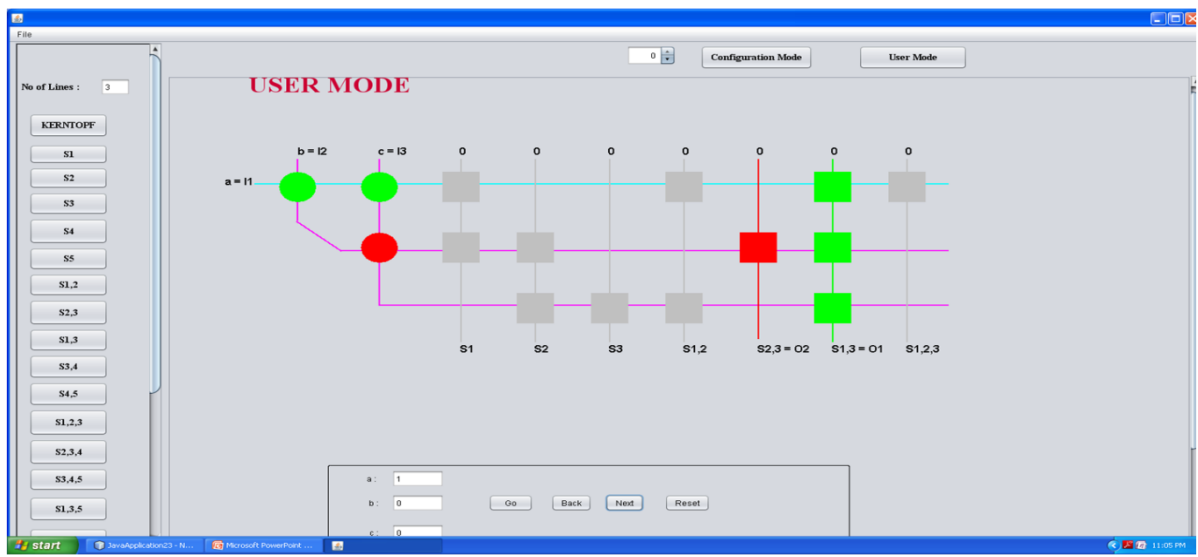

**Fig. 3.9. User Mode**

# USAGE AND RESULTS

This chapter describes the usage of the tool and results obtained. Usage part discusses and implementation session for a typical circuit. The circuit is taken from benchmark suite managed by the Maslov [30]. We describe GUI design and Experimental setup and results.

## 4.1 Graphical User Interface

In order to make the tool user friendly a Graphics user interface is implemented. After initial introduction screen two major screens namely design entry screen and RPGA screen are implemented.

### 4.1.1 Design Entry Screen

This screen caters to the need of design entry, truth table generator and symmetry analyzer. The Fig. 4.1 shows design entry screen.

Following components are used in design entry screen :

A : Drawing area

B : Selecting number of lines

C : Selecting Gate library

D : Selecting time slots

E : Putting gates on lines in latest time slot

F : Selecting Truth Table Generator

G: Selecting Symmetry Analyzer

H: Selecting RPGA Structure Entry

These components are shown in below figure. These are used to design a reversible circuit for implementation. Design Entry screen also shows the Truth Table generator for generation of irreversible truth table, symmetry analyzer for symmetry analysis of truth table and RPGA structure entry for design a RPGA structure. These are also shown in figure.

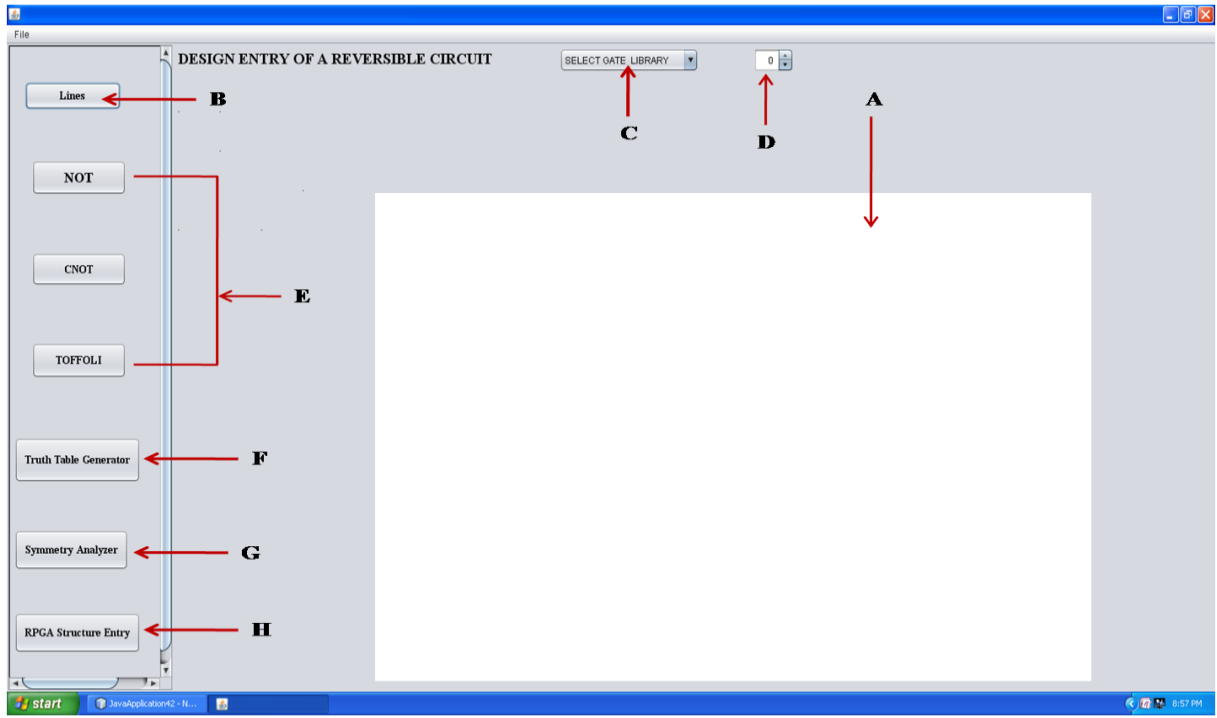

Fig. 4.1. GUI Interface for Design Entry of a Reversible Circuit

In Design Entry screen A indicates drawing area for designing a circuit. First User has to select number of lines in design screen. B indicates selecting number of lines. User can also select a gate library that is indicated by C in Fig. 4.1. Now, select time slots that is also indicated by D. Reversible gates are drawn in the respective time slot. Now, user can select reversible gates for entry in simulator. E indicates putting gates on lines in latest time slot. In design entry screen there is also some more features. Here, F indicates Truth Table Generator, G indicates Symmetry Analyzer and H indicates RPGA structure Entry. User can select F, G, H for entering respective screens.

#### 4.1.1.1 Truth Table Generator

Fig. 4.2 shows the truth table generator screen. The screen shows the irreversible truth table of a displayed circuit after design entry. User can select generate truth table shows by A for generating truth table of circuit.

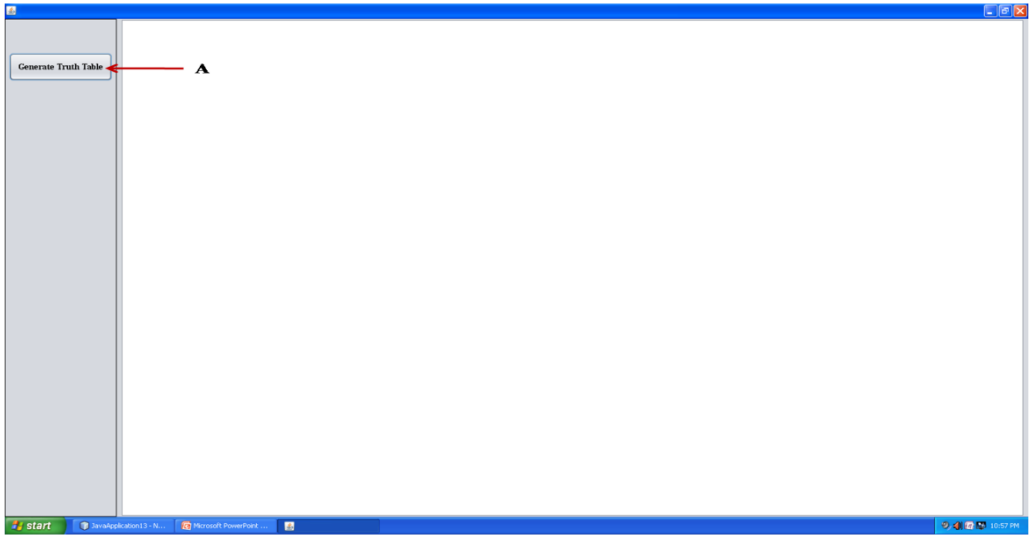

**Fig. 4.2. GUI Interface for Truth Table Generator**

### 4.1.1.2 Symmetry Analyzer

As the RPGA structure capable of simulating only symmetric circuits, analysis of the symmetry in truth table is done by symmetry analyzer in RPGA simulator. Fig. 4.3 shows symmetry analyzer screen in which user can check symmetry analysis of irreversible truth table, indicated by A.

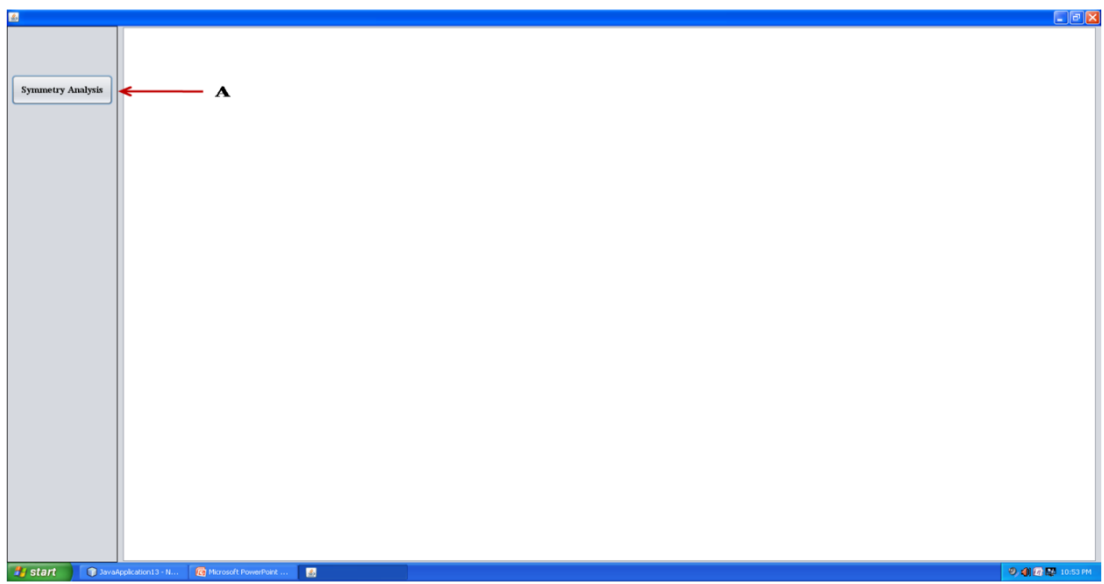

**Fig. 4.3. GUI Interface for Symmetry Analyzer**

## 4.1.2 RPGA Screen

This screen caters to the need of RPGA initial mode screen, configuration mode screen and user mode screen. Fig. 4.4 shows initial mode screen. The RPGA tool design a RPGA structure in initial mode that is displayed in initial mode screen.

In figure Components that is used in initial mode screen of RPGA as follows :

A : Drawing area

B : Selecting number of lines

C : Selecting time slots

D : Selection of RPGA structure gate (Kerntopf)

E : Selection of symmetric functions that Putting copying and feynman gates on lines in

latest time slot

F : Selecting Configuration Mode.

G: Selecting User Mode

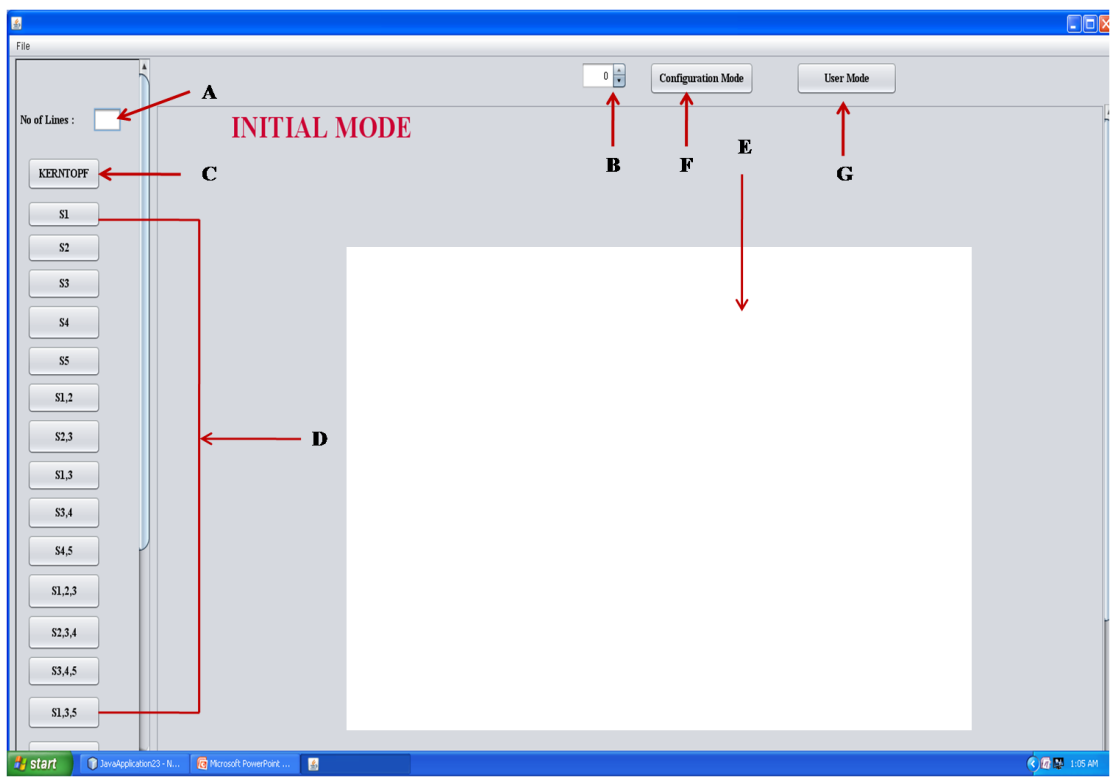

**Fig. 4.4. GUI Interface for RPGA Initial Mode Screen**

In this initial mode screen A indicates selecting number of lines. So, user has to select number of input lines to draw RPGA structure. After this, B indicates selecting time slot user can select time slots. User can select RPGA structure gates and symmetric functions that putting copying and feynman gates on lines in latest time slot. In this screen, C indicates selection of RPGA structure gate. D indicates selection of symmetric function that putting copying and Feynman gates on lines. E indicates Drawing area, that shows the RPGA structure. User can also select configuration and user mode from this screen. For this F indicates selecting configuration mode and G indicates selecting user mode.

### 4.1.2.1 Configuration Mode Screen

Fig. 4.5 shows the Configuration mode screen of RPGA. In this screen configured RPGA structure is displayed after the design RPGA structure in initial mode. Here, A indicates area in which configured RPGA structure is displayed. B indicates User mode which is used to enter in user mode screen.

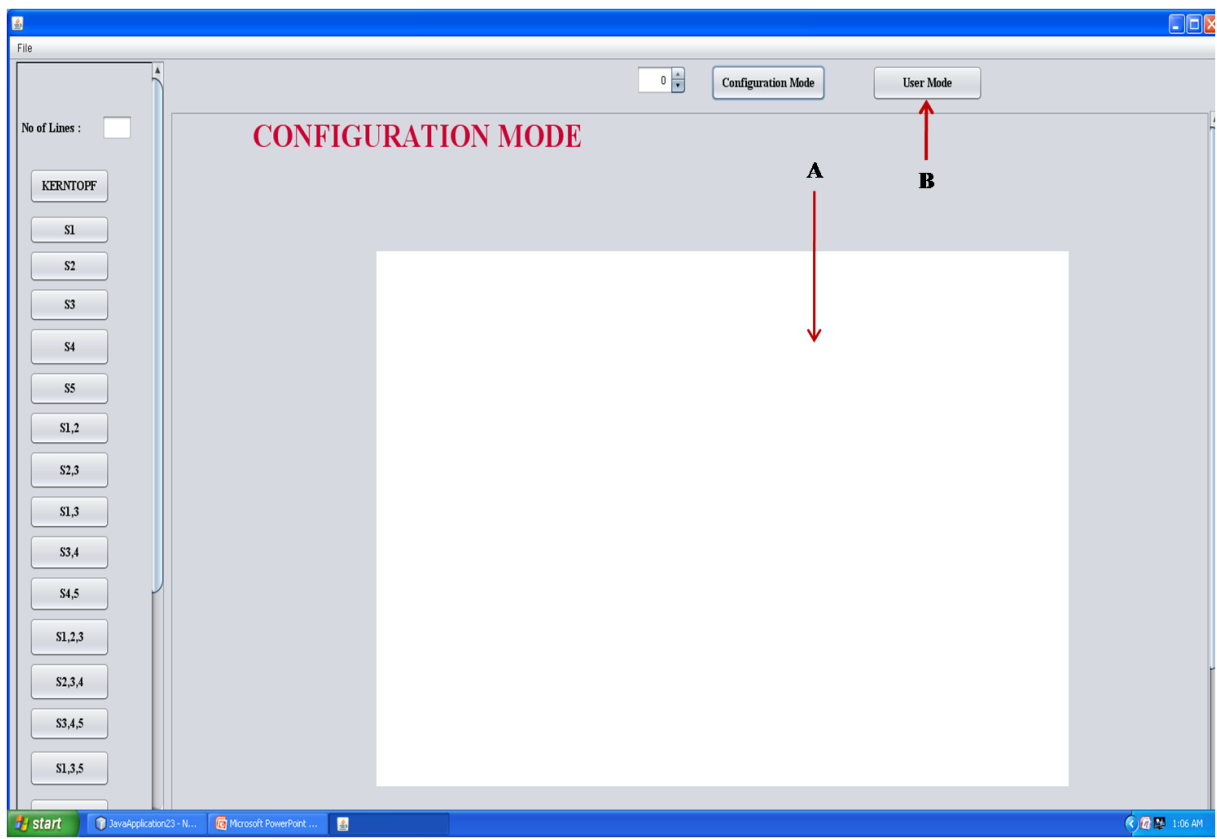

**Fig. 4.5. GUI Interface for RPGA Configuration Mode Screen**

### 4.1.2.2 User Mode screen

Fig. 4.6 shows User mode screen of RPGA. User can apply desired inputs and get the output in this user mode screen.

In figure Components that is used in User mode screen of RPGA as follows :

A : Drawing area

B : Enter user inputs

C : Selecting user inputs

D : Selecting previos user inputs

E : Selecting Next user inputs

F : Selecting Reset

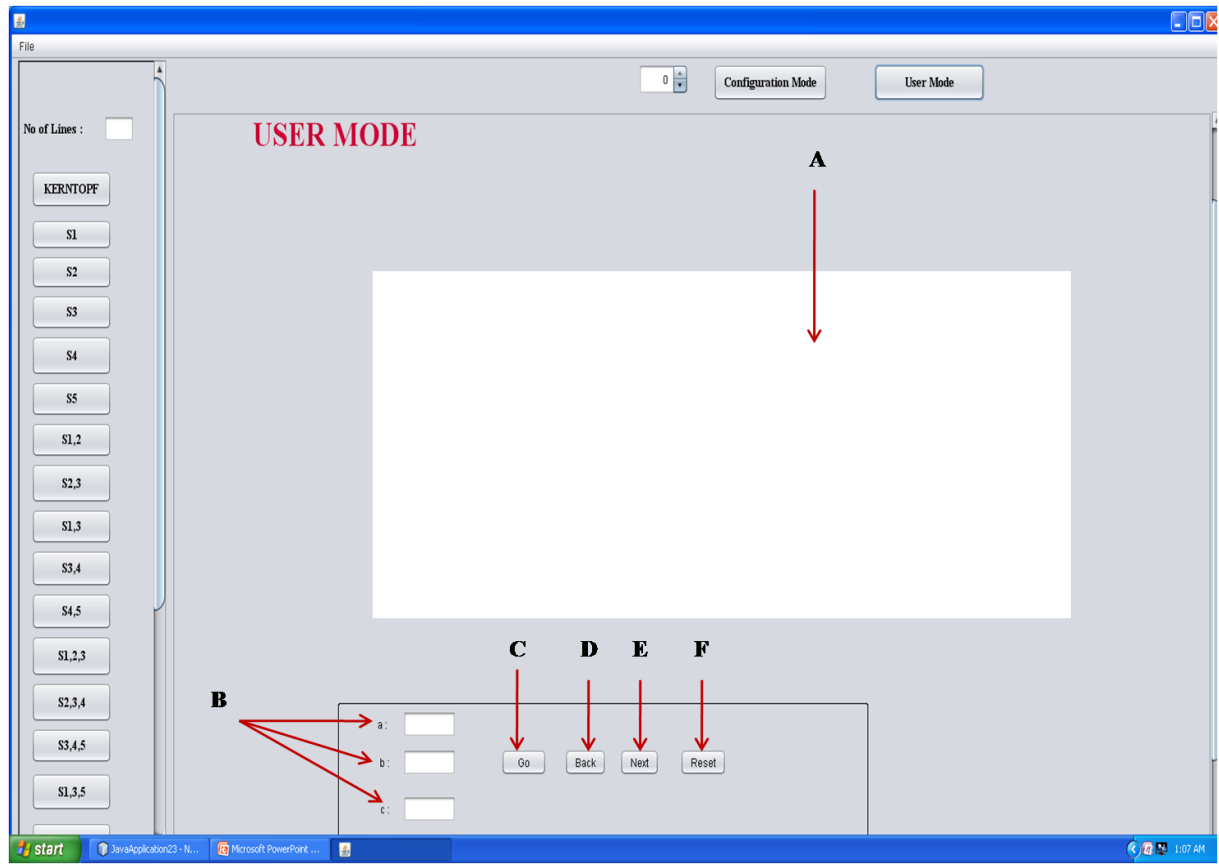

Fig. 4.6. GUI Interface for RPGA User Mode Screen

In Fig. 4.6 A indicates drawing area in which configured RPGA structure is displayed. B indicates entering user inputs of truth table. C, D and E indicates selecting user inputs, previous and next user inputs. The user can test the output for all possible inputs through next/previous button. F indicates Reset button that is used to enter again in initial mode screen.

This GUI also capable of saving RPGA structure, inclemently constructing RPGA and saving circuit and configured RPGA also.

## 4.2 Experimental setup

To test and verify the simulator experimental setup for RPGA simulator requires hardware and operating environment, benchmark suites for testing described in the following subsection.

A typical user session has also be explained in this section.

### 4.2.1 Hardware and operating environment

The tool has been developed in Java platform. We have tested the tool on the following hardware platform.

Operating System : Windows XP/7.

RAM : 512 MB, 2 GB.

JDK 1.7.0.

Netbeans IDE 7.1.2.

The tool will work on any machine with same or higher configuration and higher version of JDK/Netbeans.

### 4.2.2  Benchmark Suites

Reversible circuits are being developed by different researchers and available mostly on benchmark suite managed by the Maslov [30]. Only one suite for benchmark circuit is available. We have taken some circuits from that suite. Fig. 4.7 shows two symmetric benchmark circuit has available at that benchmark suite. We have also tested our tool on arbitrary circuit too.

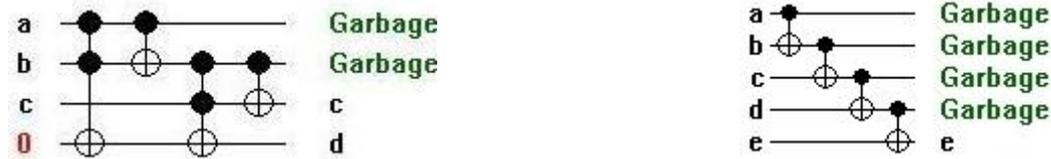

**Fig. 4.7. Symmetric Benchmark Circuit**

### 4.2.3 User Session

This section describes the step to be used for a typical implementation of circuit. Suppose a user wishes to implement a circuit from benchmark suite.

To implement a circuit on a RPGA structure following steps are used in RPGA simulator.

• First user has to enter in design entry screen after initial introduction screen.

• User has to select number of lines in this screen to draw a schematic of a reversible circuit.

• After this user can select gate library.

• User has to select time slot.

• Reversible logic gates can be drawn in respective time slots.

• After completing schematic of reversible circuit.

• Simulator generates irreversible truth table of circuit in truth table generator.

• Now symmetry analysis is done in symmetry analyzer.

• User has to developed a RPGA structure in RPGA initial mode.

• User can also enter in configured mode.

• In RPGA user mode user can apply possible inputs of truth table and get outputs.

In the Fig. 4.8 we can see a reversible benchmark circuit that is schematically design in design entry of a reversible circuit. The entire circuit is divided in time slot in such a way so that on any line there should be at most one gate in one slot.

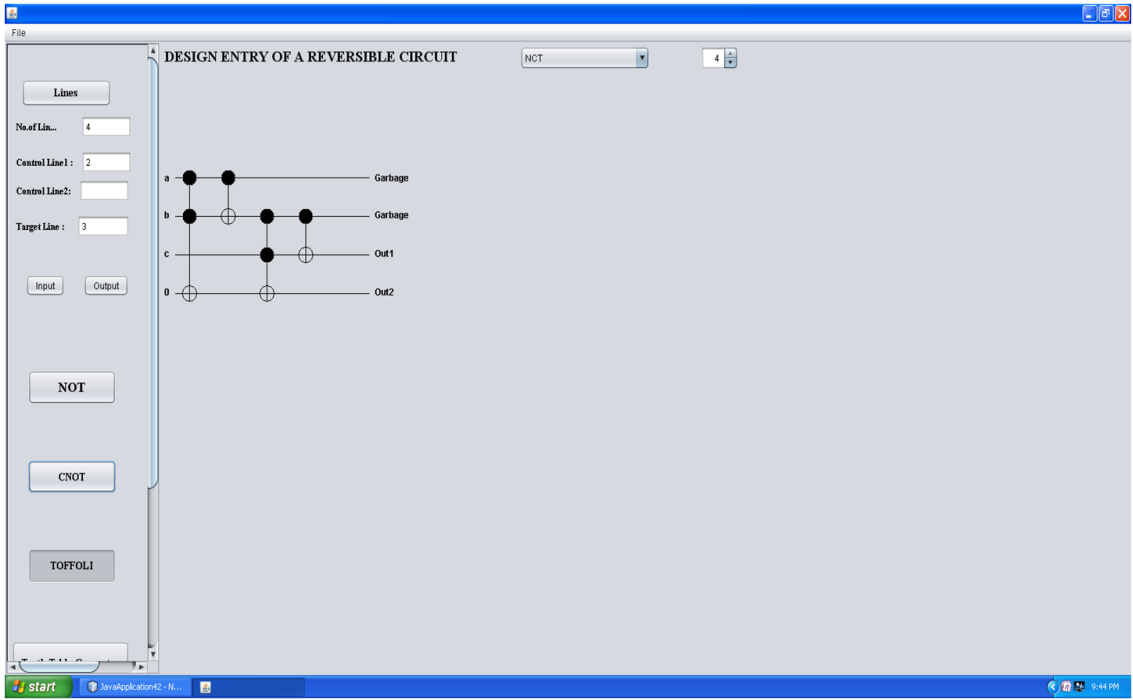

**Fig. 4.8. Reversible Circuit**

After design entry of a reversible circuit in simulator. Truth table of a reversible circuit is generated. The truth table generator generates reversible truth table followed by irreversible truth table. So, user can view the irreversible truth table in truth table generator screen. In Fig. 4.9 there is irreversible truth table of reversible 1bitadder (r32) circuit.

```
Input    Output
0 0 0    0 0
0 0 1    1 0
0 1 0    1 0
0 1 1    0 1
1 0 0    1 0
1 0 1    0 1
1 1 0    0 1
1 1 1    1 1
```

**Fig. 4.9. Generated Truth Table of Reversible Circuit**

As we already described that RPGA structure capable of simulating only symmetric circuits, analysis of the symmetry in truth table of the circuit is necessary. User can check symmetry analysis in symmetry analyzer.

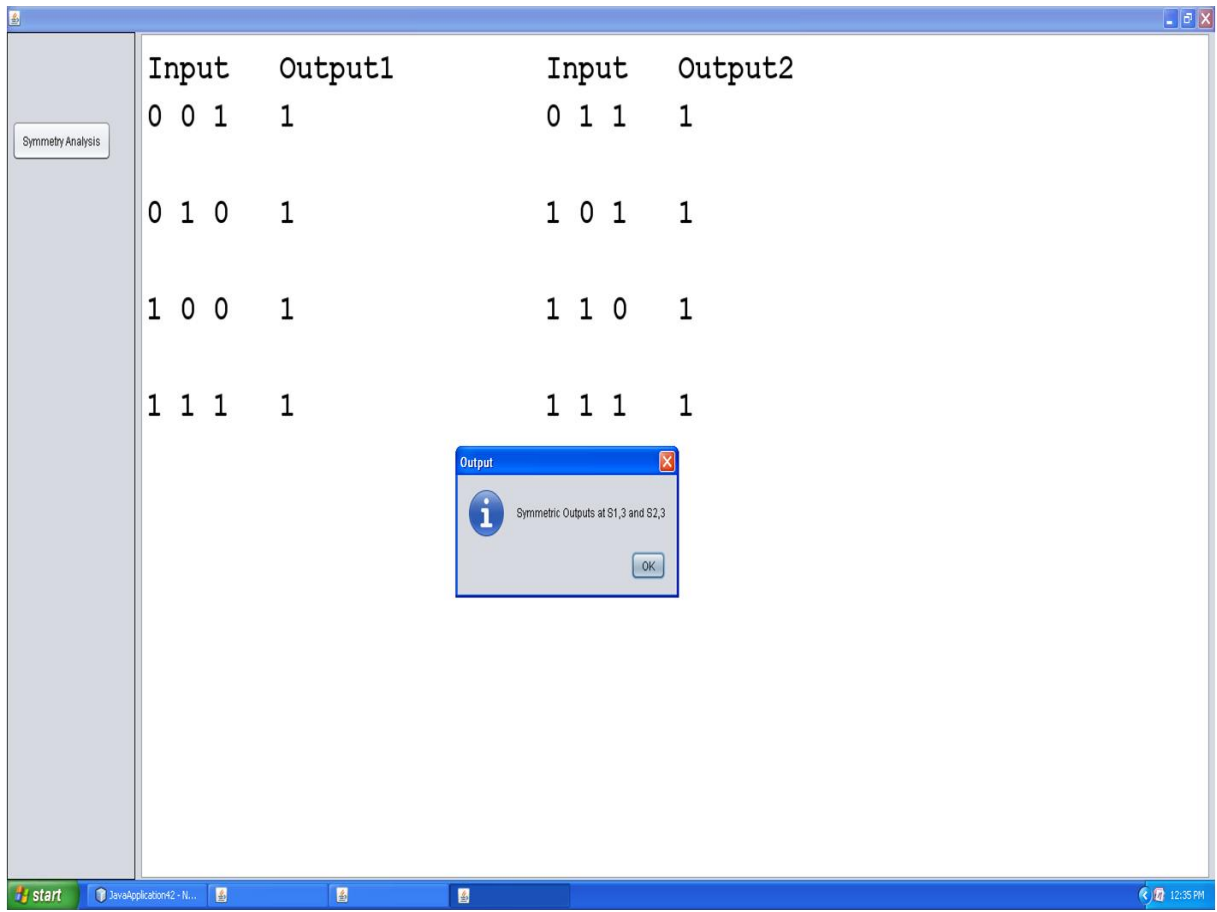

**Fig. 4.10. Symmetry Analysis of generated truth table**

In this part Simulator takes no of lines as a input and schematically create a RPGA structure using RPGA structure gates and other copying and Feynman gates. Here, we create a RPGA structure for 3 inputs. User can view the RPGA structure in initial mode screen of RPGA.

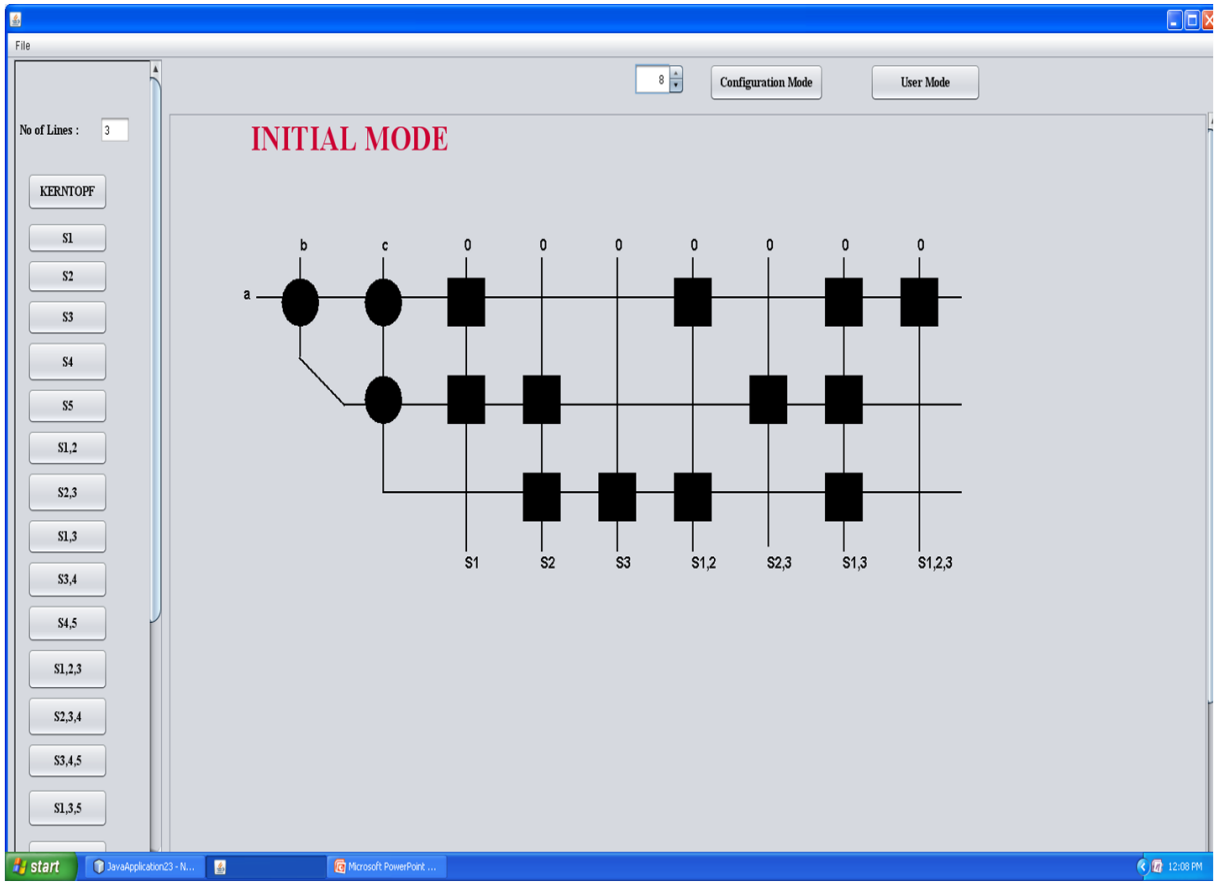

**Fig. 4.11. RPGA structure**

After this step circuit is to be implemented on RPGA structure. Next section shows the implementation results.

### 4.2.4 Results

The RPGA simulator has been tested and the circuit implementation has been verified for the following benchmark suite circuit.

Reversible Benchmark Circuit : 1bitadder(r32).

Circuit has been implemented on a RPGA structure in RPGA simulator. Fig. 4.12 shows the result at the particular user inputs. Suppose user apply input I1 = 1, I2 = 0 AND I3 = 0. User can get the result at S1,3 for output1 and S2,3 for output2. So, for I1 =1, I2 = 0 I3 = 0 green color indicates on at S1,3 = O1 and red color indicates off at S2,3 = O2.

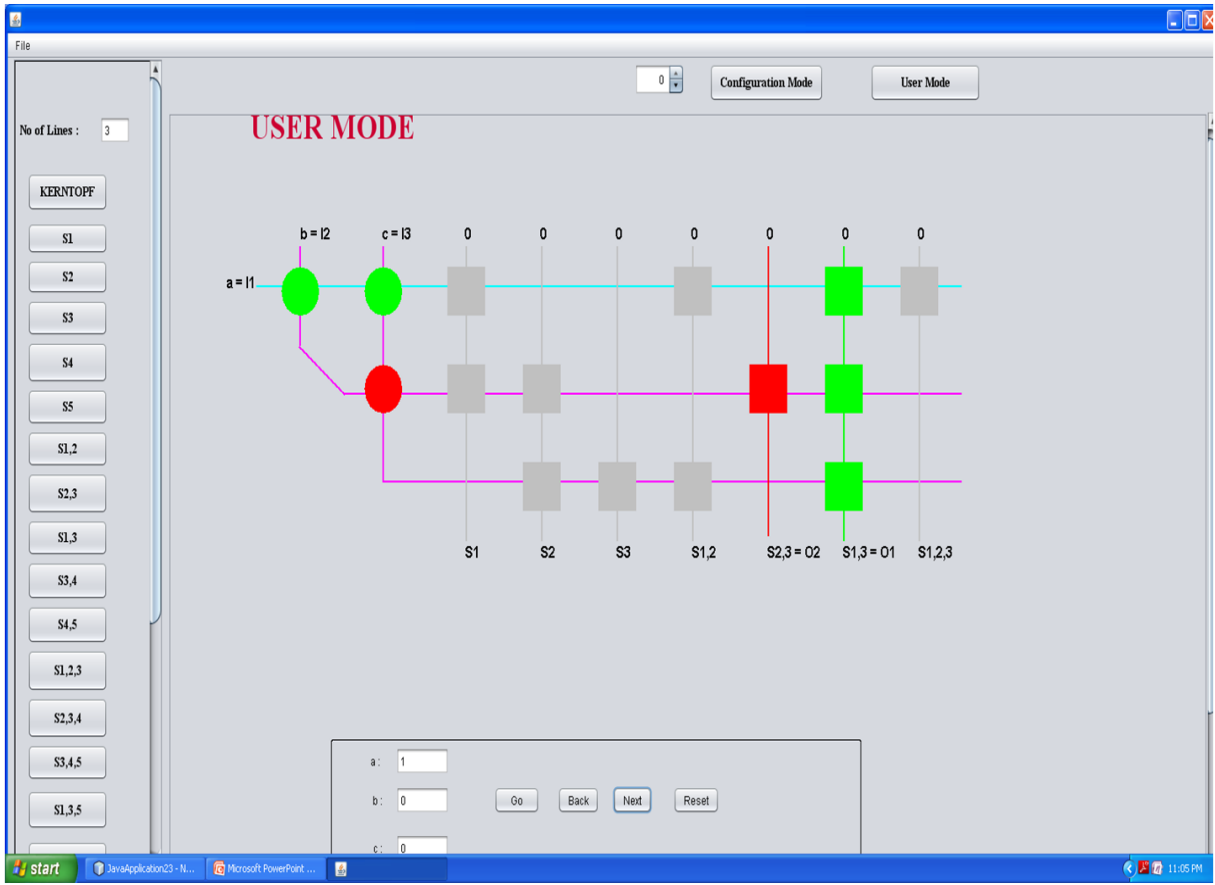

**Fig. 4.12. Implementation Results**

Hence we observed that the developed simulator is working well and can be used testing any circuit.

# CONCLUSION AND FUTURE SCOPE

## 5.1 Conclusion

In our dissertation work, we have undertaken the development of a RPGA simulator for implementing and testing reversible circuits. The simulator is also capable of implementing irreversible circuits on RPGA. As no platform is available as of now for implementation of reversible circuits including RPGA, RPGA simulator will be of immense use for designers working in this area.

Our specific contributions in this dissertation are as under.

• Design entry of reversible circuits from schematic.

• Editing, saving and retrieval stored reversible circuits.

• Reversible and irreversible truth table generation.

• Symmetry analyzer.

• RPGA structure entry.

• Implementation of simulation environment.

Hence our tool provides a complete solution for reversible as well as irreversible circuits. This has filled a great need of designers.

## 5.2 Future Scope

The RPGA structures proposed in literature are able to implement the behaviour of symmetric functions/circuits only. A new structure technique is needed to implement asymmetric functions also. Presently Picton and kerntopf gates are used for MAX/MIN function. A better structure with less garbage is also required to reduce the implementation complexity. A large number of copying gate plane lines are required in present RPGA structures. The reduction in such lines is required in improved RPGA structures. Presently binary valued circuits are implemented in proposed RPGA structure should implement multivalued circuits too.

# BIBLIOGRAPHY

[1] R. Landauer, "Irreversibility and heat generation in the computational process", I.B.M. Journal of Research and Development, 5 (1961), pp. 183-191.

[2] Bennet C., "Logical Reversibility of Computation ", IBM Journal of Research and Development, vol. 17, no. 6, pp. 525-532, 1973.

[3] Shazia Hassan, "Limitation Of Silicon Based Computation And Future Prospects", Second International Conference on Communication Software and Networks, IEEE , 2010.

[4] Michael P. Frank, "Physical Limits of Computing", IEEE Computing in Science & Engineering magazine, May/June 2002.

[5] Ch. H. Bennett and R. Landauer, "The Fundamental Limits of Computation", Scientific American, July 1985, pp. 38-46.

[6] Rangaraju H G, Venugopal U, Muralidhara K N, Raja K B, "Low Power Reversible Parallel Binary Adder/Subtractor", International journal of VLSI design & Communication Systems (VLSICS) Vol.1, No.3, September 2010.

[7] Ch.H. Bennett , "Time/Space Trade-offs For Reversible Computation", IBM J. Res. Dev., SIAM J. COMPUT. Vol. 18, No. 4, pp. 766-776, August 1989.

[8] Michael P. Frank, "Introduction to Reversible Computing: Motivation, Progress, and Challenges", CF'05, May 4–6, 2005, Ischia, Italy.

[9] Vivek V. Shende, Aditya K. Prasad, Igor L. Markov, and John P. Hayes, "Synthesis of Reversible Logic Circuits", IEEE TRANSACTIONS ON COMPUTER-AIDED DESIGN OF INTEGRATED CIRCUITS AND SYSTEMS, VOL. 22, NO. 6, JUNE 2003.

[10] N.Srinivasa Rao,P.Satyanarayana "A Novel Reversible Gate and Application", International Journal of Engineering and Technology Volume 2 No 7 ,July ,2012.

[11] Mehdi Saeedi, Igor L. Markov, "Synthesis and Optimization of Reversible Circuits - A Survey", 2011.

[12] Y. Kanamori, S.-M. Yoo, W.D. Pan and F.T. Sheldon "A SHORT SURVEY ON QUANTUM COMPUTERS", International Journal of Computers and Applications, Vol. 28, No. 3, 2006

[13] Soo-Hong Kim and Sung Choi, "Scalable Systolic Structure to Realize Arbitrary Reversible Symmetric Functions", GESTS Int'l Trans. Computer Science and Engr., Vol.18, No.1, 2005.

[14] Marek Perkowski, Pawel Kerntopf, Andrzej Buller, Malgorzata Chrzanowska-Jeske, Alan Mishchenko, Xiaoyu Song, Anas Al-Rabadi, Lech Jozwiak, Alan Coppola, Bart Massey, "Regular Realization of Symmetric Functions Using Reversible Logic", IEEE 2001.

[15] Raghava Garipelly, P.Madhu Kiran, A.Santhosh Kumar, "A Review on Reversible Logic Gates and their Implementation", International Journal of Emerging Technology and Advanced Engineering, SSN 2250 - 2459 Volume 3 , Issue 3 , March 2013

[16] Prashant .R.Yelekar, Prof. Sujata S. Chiwande, "Introduction to Reversible Logic Gates & its Application", 2nd National Conference on Information and Communication Technology (NCICT) 2011.

[17] B.Raghu kanth, B.Murali Krishna, M. Sridhar, V.G. Santhi Swaroop "A DISTINGUISH BETWEEN REVERSIBLE AND CONVENTIONAL LOGIC GATES", International Journal of Engineering Research and Applications (IJERA) Vol. 2, Issue 2,Mar-Apr

2012, pp.148-151.

[18] Ravish Aradhya H V, Praveen Kumar B V, Muralidhara K N, "Design of Control unit for Low Power ALU Using Reversible Logic", International Journal of Scientific & Engineering Research Volume 2, Issue 9, September-2011

[19] Himanshu Thapliyal, M.B Srinivas "A New Reversible TSG Gate and Its Application For Designing Efficient Adder Circuits"

[20] T. Toffoli, - Reversible Computing, MIT Lab for Computer Science, Technical memo MIT/LCS/TM-151, Feb 1980.

[21] Toffoli "Reversible Computing", in Automata, Languages and Programming, Springer-Verlag, 1980, pp. 632-644.

[22] Dmitri Maslov, Gerhard W. Dueck, and D. Michael Miller, "Synthesis of Fredkin–Toffoli Reversible Networks", 1063-8210, 2005 IEEE

[23] Majid Haghparast and Keivan Navi, "A Novel Fault Tolerant Reversible Gate For Nanotechnology Based Systems", American Journal of Applied Sciences 5 (5): 519-523, 2008.

[24] Saiful Islam, Muhammad Mahbubur Rahman, Zerina Begum, Mohd Zulfiquar, "Synthesis of Fault Tolerant Reversible Logic Circuits", IEEE Journal, Publication Year : 2009.

[25] Behrooz Parhami, "Fault-Tolerant Reversible Circuits", Proc. 40th Asilomar Conf. Signals, Systems, and Computers, Pacific Grove, CA, October 2006.

[26] Fredkin E., "Conservative Logic", International Journal of Theoretical Physics, pp. 219- 253, 1982.

[27] Giovanni De Micheli, "Synthesis And Optimization Of Digital Circuits" , TATA McGraw- Hill Edition.

[28] M. Perkowski, P. Kerntopf, A. Buller, M. Chrzanowska-Jeske, A. Mishchenko, X. Song, A. Al-Rabadi, L. Jozwiak, A. Coppola, "Regular Realization of Symmetric Functions using Reversible Logic", submitted to Euro-Micro 2001.

[29] M.Perkowski, A.Al-Rabadi, P.Kerntopf, A.Mishchenko and M.Chrzanowska- Jeske, A.Buller, A.Coppola, and L.Jozwiak, "Regularity and Symmetry as a Base for Efficient Re-alization of Reversible Logic Circuits", Proc. 10-th International Workshop on Logic and Syn-thesis, IWLS'01 , pp. 90 - 95, 2001.

[30] Dmitri Maslov , "Reversible Logic Synthesis Benchmarks circuits" .